%% file: main.tex
\PassOptionsToPackage{hyphens}{url}
\documentclass[sigconf, screen, nonacm]{acmart}

\renewcommand\footnotetextcopyrightpermission[1]{}
\AtBeginDocument{%
  }

\usepackage{amsmath}
\usepackage{algorithmic}
\usepackage{graphicx}
\usepackage{textcomp}
\usepackage{xcolor}
\usepackage{xspace}
\usepackage{makecell}
\usepackage{pifont}
\usepackage{cleveref}
\usepackage{booktabs}
\usepackage[most]{tcolorbox}
\usepackage{tikz}

\newcommand*\blackcircle[1]{%
  \tikz[baseline=(char.base)]{
    \node[
      shape=circle,
      fill=black,
      inner sep=0.4pt,
      minimum size=8pt,
      text=white,
      font=\sffamily\bfseries\footnotesize
    ] (char) {#1};
  }%
}

\tcbset{
  insight/.style={
    enhanced,
    colback=gray!2,
    colframe=black!35,
    boxrule=0.3pt,
    arc=2pt,
    left=6pt, right=6pt, top=4pt, bottom=4pt,
    before skip=6pt, after skip=6pt,
    fontupper=\itshape,
    sharp corners=all,
  }
}

\newtcolorbox{mybasecolorbox}[1][]{%
 colback=gray!10, colframe=black,
 coltitle=black, fonttitle=\small\bfseries,
 sharp corners,
 width=1.0\linewidth,
 title=#1,
 left=1pt,
 right=1pt,
 top=1pt,
 bottom=1pt,
 boxsep=2pt,
 boxrule=0.5pt,
 before=\vspace{1pt},
 after=\vspace{1pt},
 }

\definecolor{mygray}{RGB}{235,235,235}

\usepackage[switch]{lineno}
\begin{document}

\newcommand{\moepim}{\textsc{Sieve}\xspace}
\newcommand{\runtime}{\textsc{xPU-Runtime}\xspace}
\newcommand{\pimdevice}{devices}
\newcommand{\todo}[1]{\textcolor{red}{TODO: #1}}
\newcommand{\circnum}[1]{\ding{\numexpr171+#1\relax}}

\title{Sieve: Dynamic Expert-Aware PIM Acceleration for Evolving Mixture-of-Experts Models}

\author{Jungwoo Kim}
\affiliation{
  \institution{Stanford University}
  \city{Stanford}
  \state{California}
  \country{USA}
}
\email{jungwkim@stanford.edu}

\author{Rubens Lacouture}
\affiliation{
  \institution{Stanford University}
  \city{Stanford}
  \state{California}
  \country{USA}
}
\email{rubensl@stanford.edu}

\author{Genghan Zhang}
\affiliation{
  \institution{Stanford University}
  \city{Stanford}
  \state{California}
  \country{USA}
}
\email{zgh23@stanford.edu}

\author{Gina Sohn}
\affiliation{
  \institution{Stanford University}
  \city{Stanford}
  \state{California}
  \country{USA}
}
\email{ginasohn@stanford.edu}

\author{Qizheng Zhang}
\affiliation{
  \institution{Stanford University}
  \city{Stanford}
  \state{California}
  \country{USA}
}
\email{qizhengz@stanford.edu}

\author{Swapnil Gandhi}
\affiliation{
  \institution{Stanford University}
  \city{Stanford}
  \state{California}
  \country{USA}
}
\email{gandhis@stanford.edu}

\author{Christos Kozyrakis}
\affiliation{
  \institution{Stanford University}
  \city{Stanford}
  \state{California}
  \country{USA}
}
\affiliation{
  \institution{NVIDIA}
  \city{Santa Clara}
  \state{California}
  \country{USA}
}
\email{kozyraki@stanford.edu}

\author{Kunle Olukotun}
\affiliation{
  \institution{Stanford University}
  \city{Stanford}
  \state{California}
  \country{USA}
}
\email{kunle@stanford.edu}

\renewcommand{\shortauthors}{Jungwoo Kim et al.}

\input{sections/0_abstract}


\maketitle

\input{sections/1_introduction}
\input{sections/2_background}
\input{sections/3_motivation}
\input{sections/4_overview}
\input{sections/5_scheduler}
\input{sections/6_system}
\input{sections/7_evaluation}
\input{sections/8_related}

\input{sections/9_conclusion}

\bibliographystyle{ACM-Reference-Format}
\bibliography{refs}

\end{document}

%% file: sections/0_abstract.tex
\begin{abstract}

Mixture-of-Experts (MoE) has become a dominant architecture for scaling large language models (LLMs).
However, the execution characteristics of MoE inference are changing rapidly and increasingly mismatch the assumptions underlying existing Processing-in-Memory (PIM) systems.
Prior PIM systems for LLMs rely on static rules to offload memory-bound operations to PIM, without accounting for the combined effects of load imbalance and inter-GPU communication.
Meanwhile, modern MoE models activate fewer experts out of increasingly many, creating a bimodal expert distribution: a small set of experts receives many tokens, while a long tail of experts receives only one or a few.

We identify a trend in modern MoE models toward increasingly bimodal token-to-expert distributions, quantify the resulting disparity in arithmetic intensity across experts, and show that this disparity dramatically reduces the efficiency of state-of-the-art PIM systems for LLMs.
To address this problem, we propose a scheduler for serving MoE models on multi-GPU systems with attached HBM-PIM stacks. 
Our scheduler partitions expert execution between GPU and PIM based on runtime token-to-expert distributions, while jointly considering interconnect overhead, memory bandwidth, GPU throughput, and PIM throughput.
Moreover, we propose \moepim, a runtime framework that employs the scheduler to coordinate execution across GPUs and their attached HBM-PIM stacks. \moepim overlaps GPU computation, PIM computation,
and intra- and inter-device communication while preserving cross-device dependencies induced by expert parallelism.
\moepim is evaluated on our cycle-accurate simulator based on Ramulator 2.0.
Compared to state-of-the-art PIM systems for MoE, \moepim improves both throughput and interactivity by $1.3\times$, $1.3\times$, and $1.6\times$ on Qwen3.5-397B-A17B, GPT-OSS-120B, and Qwen3-30B-A3B, respectively.
\end{abstract}

%% file: sections/1_introduction.tex
\section{Introduction}

The Mixture-of-Experts (MoE) architecture has emerged as a leading direction for scaling LLM capacity efficiently~\cite{shazeer2017outrageously, fedus2022switch,du2022glam, lepikhin2020gshard}.
Instead of processing every token through a single dense feed-forward network (FFN), MoE models activate only a small subset of \textit{experts} per token, dramatically increasing total parameter capacity without a proportional increase in per-token computation~\cite{shazeer2017outrageously, fedus2022switch,du2022glam, lepikhin2020gshard}.
State-of-the-art LLMs are increasingly adopting MoE layers to 
improve specialization and efficiency, and reduce training and inference cost, 
making MoE central to the next generation of LLMs~\cite{gemini2.5,qwen3,gptoss,kimiK2,deepseekR1,deepseekV3,glm4.5}.

At the same time, serving these increasingly large models is becoming bottlenecked not by arithmetic throughput, but by the cost of moving billions of parameters through the memory hierarchy~\cite{dally2023hardware}. 
This growing imbalance, known as the AI memory wall~\cite{aiandmemorywall}, has motivated the adoption of Processing-in-Memory (PIM) technology.
PIM architectures embed lightweight compute units near DRAM banks to exploit the high internal bandwidth of modern memory devices~\cite{samsung_hbm2pim,hynix_pim_isscc22}.
Prior work shows that PIM can substantially accelerate memory-bound components of LLM inference, such as attention layers~\cite{paise,papi,duplex,neupims,attacc}.

\begin{figure}
    \centering
    \includegraphics[width=\columnwidth]{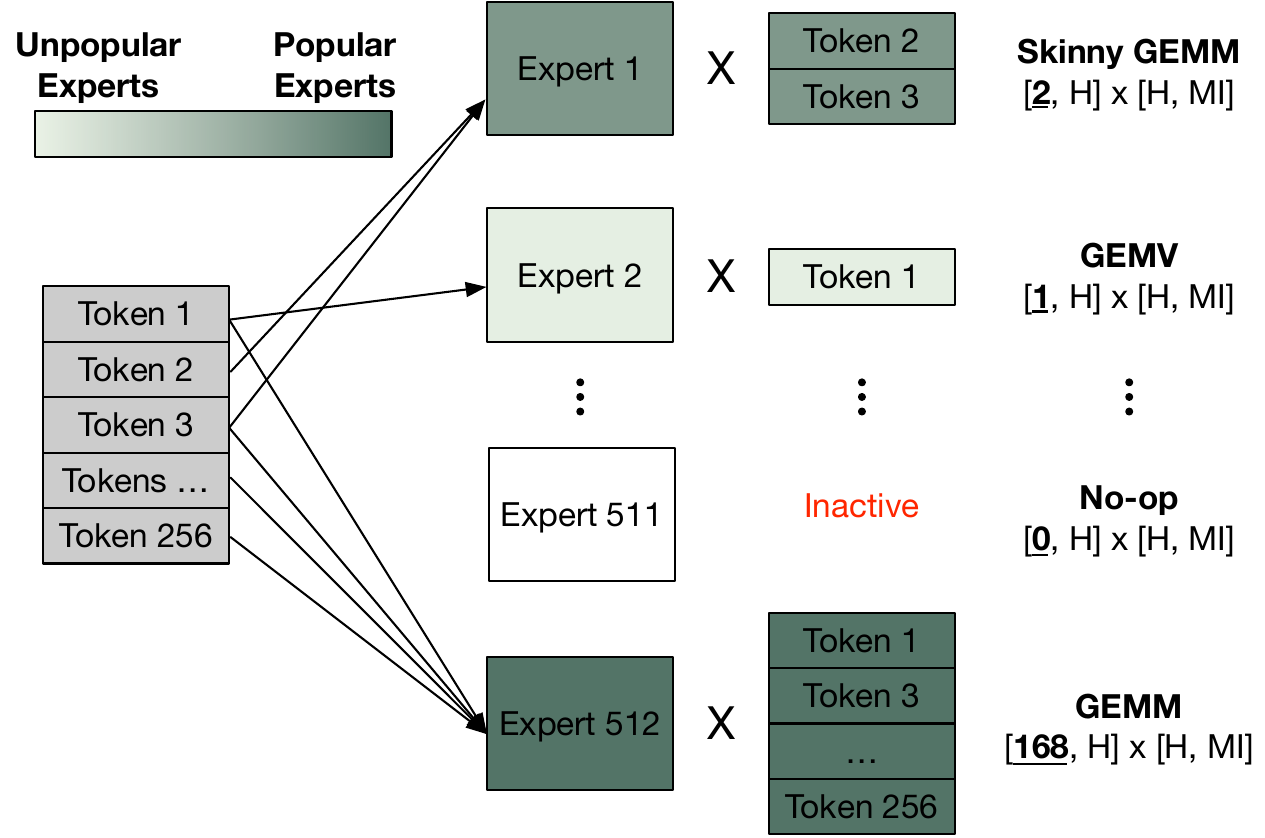}
    \Description{An example expert distribution in which token allocation varies across experts.}
    \caption{An example expert distribution in which token allocation varies across experts.
    }
    \label{fig:expert-distribution}
\end{figure}

However, modern MoE models are evolving in ways that fundamentally challenge existing PIM-enabled LLM systems.
As illustrated in \Cref{fig:expert-distribution}, fewer experts are activated out of increasingly many~\cite{gptoss, kimiK2, glm4.5, deepseekV3, qwen3}, yielding a bimodal token-to-expert distribution where a small set of popular experts receives many tokens while a long tail of unpopular experts receives only one or a few. 
This distribution creates a large disparity in arithmetic intensity across experts.
For example, in \textit{Qwen3-Next} at batch size 64,
44.2\% of experts receive only a single token, and 89.3\% receive at most four, leaving only a small minority to process larger token batches.

We further show that this disparity makes state-of-the-art PIM systems inefficient. 
Existing PIM systems rely on \textbf{static} offloading rules, such as mapping only attention to PIM or using a fixed threshold to determine whether an expert should run on PIM~\cite{neupims,paise, pimoe}.
Prior work also assumes a global interconnect across PIM devices, overlooking the combined effects of expert imbalance and inter-GPU communication that commonly arise when serving MoE models in multi-GPU systems~\cite{neupims, attacc, papi, pimoe}.

Building on this analysis, we propose a scheduler for serving MoE models on multi-GPU systems with HBM-PIM stacks.
The scheduler exploits the arithmetic intensity disparity induced by the bimodal expert distribution to \textbf{dynamically} partition expert computation between GPUs and their attached HBM-PIM stacks.
In general, the scheduler dispatches experts with low arithmetic intensity to PIM, and those with high arithmetic intensity to GPUs.

However, naively assigning memory-bound experts to PIM is suboptimal because expert parallelism across GPUs introduces inter-GPU communication, and the attention operations must co-execute with memory-bound experts on PIM.
To improve the performance, our scheduler jointly accounts for interconnect overhead, memory bandwidth, GPU throughput, and PIM throughput.
This differs from prior work, which neglects the all-to-all communication overhead and assumes that a GPU can access any PIM device~\cite{neupims, papi, pimoe}.
Moreover, our scheduler uses a lightweight runtime algorithm, incurring only 20µs overhead on an NVIDIA B200 GPU.

We also design \moepim, a runtime framework that incorporates the new scheduler to enable practical deployment and efficient coordination of multi-GPU systems with HBM-PIM stacks.
It coordinates execution across GPUs and their attached HBM-PIM stacks by overlapping PIM computation, GPU computation, and inter-GPU communication to maximize hardware utilization.
In addition, \moepim employs expert parallelism across GPUs and tensor parallelism across the PIM channels attached to each GPU, thereby maximizing PIM utilization by ensuring that no PIM channel is left underutilized when memory-bound experts are assigned to PIM.
It also preserves efficient GPU execution for popular experts through grouped GEMM, even when some tokens originally dispatched to the GPU are offloaded to PIM, requiring additional aggregation between the GPU and its attached PIM.

We evaluate \moepim using three state-of-the-art MoE models with different sparsity patterns, model sizes and ratios of activated to total parameters: GPT-OSS-120B (\textit{GPT-OSS}), Qwen3.5-397B-A17B (\textit{Qwen3.5}), and Qwen3-30B-A3B (\textit{Qwen3}).
Compared to the state-of-the-art PIM systems for MoE, \moepim achieves $1.3\times$, $1.3\times$, and $1.6\times$ improved throughput and interactivity on \textit{Qwen3.5}, \textit{GPT-OSS}, and \textit{Qwen3}, respectively. 

In this paper, we make the following key contributions:
\begin{enumerate}
    \item We \textbf{identify and evaluate a trend in recent MoE models} toward increasingly bimodal expert distributions, which create disparities in arithmetic intensity across experts, and \textbf{show the impact of this trend} that renders state-of-the-art PIM-enabled systems inefficient.
    
    \item We propose a scheduler for serving MoE models in multi-GPU systems with PIM \textbf{that exploits the bimodal expert distribution and uses arithmetic intensity} derived from runtime token-expert distributions to dynamically partition expert computations between each GPU and its attached HBM-PIM stacks.
    
    \item We design \moepim, a runtime framework that employs the new scheduler to efficiently coordinate execution across GPUs and attached HBM-PIM stacks. \moepim \textbf{overlaps computation across GPUs and attached HBM-PIM stacks with both inter- and intra-device communication}.
    
    \item We evaluate \moepim on state-of-the-art MoE models and show that \moepim's gains are robust across batch sizes and workload distributions.
\end{enumerate}

%% file: sections/2_background.tex
\section{Background}
\label{sec:background}
The two key components of our work are Mixture-of-Experts (MoE) models and Processing-in-Memory (PIM) architectures.
We first discuss the design of MoE architectures, and then review how PIM addresses the memory bottlenecks in LLM acceleration.

\subsection{Mixture-of-Experts (MoE)}
\begin{figure}
    \centering
    \includegraphics[width=\columnwidth]{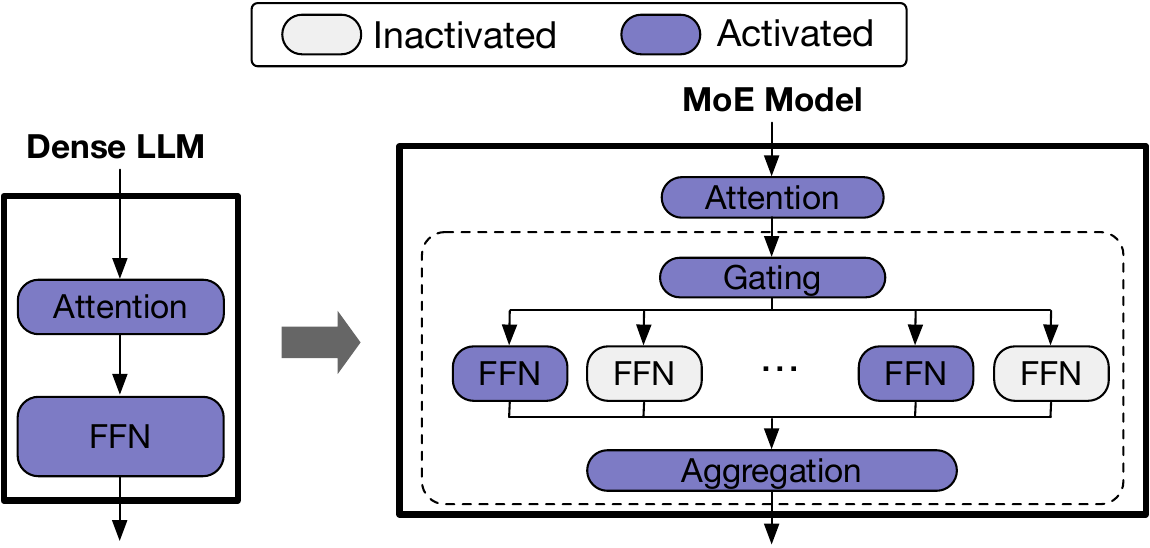}
    \Description{Comparison of the key differences between dense LLM and MoE architectures.}
    \caption{Comparison of the key differences between dense LLM and MoE architectures.}
    \label{fig:dense_vs_moe}
\end{figure}

As illustrated in Figure~\ref{fig:dense_vs_moe},
MoE models differ from dense LLMs by processing each token through a distinct combination of experts instead of a single shared FFN.
A MoE layer consists of multiple independent FFNs, referred to as ``experts'', and a gating network that determines which experts each token selects.
The selected experts process their assigned tokens, and an aggregation unit combines the corresponding outputs.
This selective activation allows MoE models to scale total parameter capacity dramatically while keeping per-token computation and memory requirements low.

\begin{figure}
    \centering
    \includegraphics[width=\columnwidth]{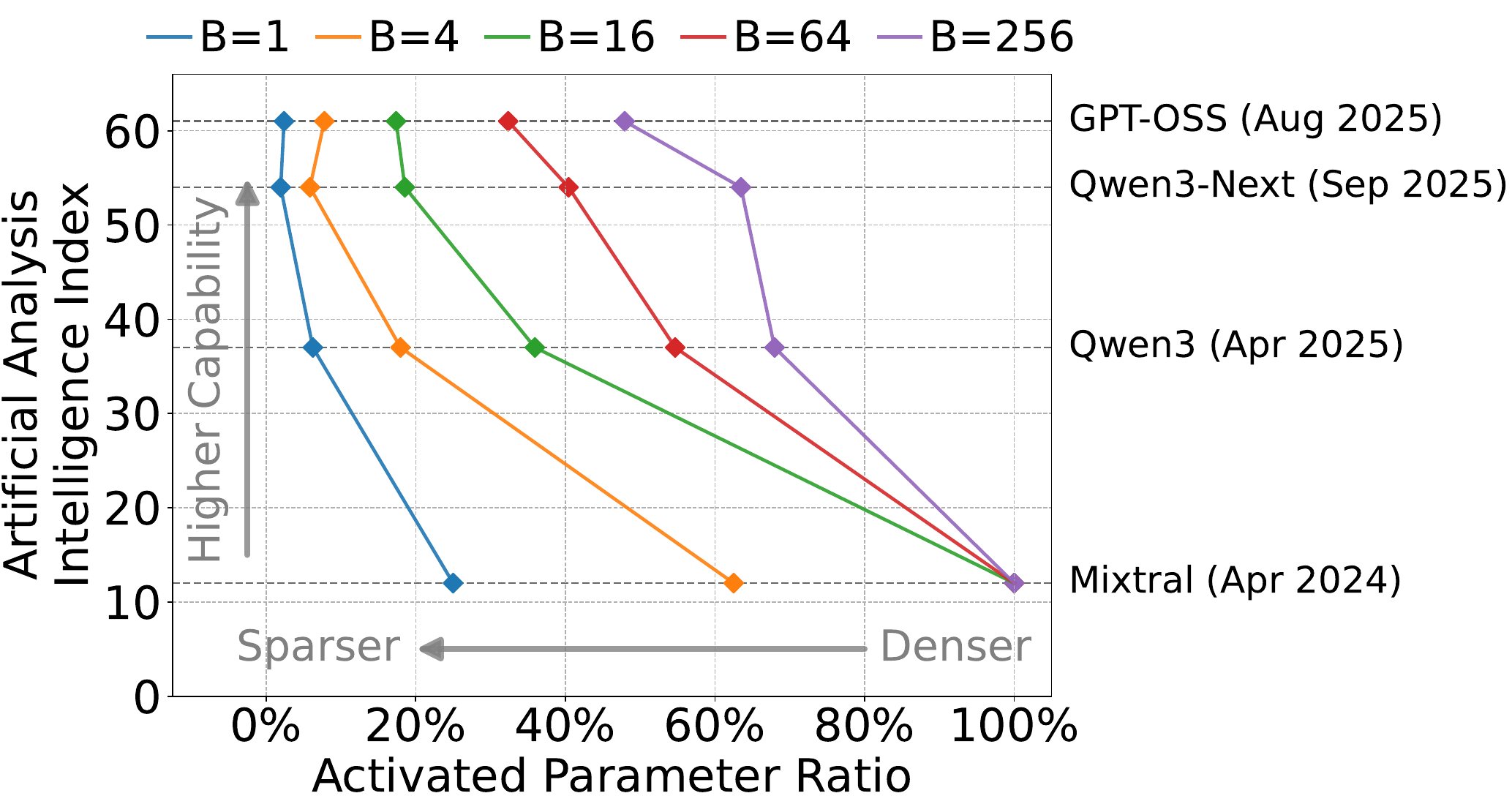}
    \Description{Benchmarking the sparsity-capability relationship showing median activated-parameter ratio versus AAII score.}
    \caption{Benchmarking the sparsity–capability relationship. The plot shows the median activated-parameter ratio, where a lower ratio indicates a sparser model.
    AAII is considered batch-invariant.
    }
    \label{fig:activated_ratio}
\end{figure}

\subsection{Processing-in-Memory for LLMs}
\label{sec:pim-for-llms}

PIM has emerged as a promising architectural solution to the memory-wall problem in modern accelerators~\cite{samsung_hbm2pim, hynix_pim_isscc22,samsungPIM_hcs23, gddr6aim, aquabolt, newton}. 
By placing processing units (PUs) near DRAM banks, PIM architectures directly leverage the large internal memory bandwidth for computation, thereby bypassing the limited external memory bandwidth of GPUs. 
Since this internal bandwidth can exceed the external bandwidth by an order of magnitude, PIM provides strong potential for accelerating memory-bound operations.
PIM architectures are particularly efficient for general matrix-vector multiplication (GEMV) operations, as their PUs are often specialized for dot product operations using adder trees~\cite{hynix_pim_isscc22, newton, samsungPIM_hcs23}. 
To exploit bank-level parallelism, the vector operand is typically broadcast to all PUs within a PIM channel, while the matrix operand is partitioned across the banks associated with the PIM channel~\cite{newton, neupims, paise}.

In the context of LLM inference, prior studies identify the attention operation in the decoding phase as memory-bound, characterized by low arithmetic intensity and dominated by GEMV operations~\cite{paise,papi,duplex,neupims,attacc,pimisallyouneed}. 
Based on this observation, PIM-enabled systems commonly offload attention operations to PIM, while compute-bound GEMM operations in FFN and QKV generation layers remain on GPUs or NPUs.
This offloading strategy yields approximately a $2\times$ speedup over PIM-disabled systems~\cite{paise,duplex,neupims,attacc,pimisallyouneed}, indicating a shift from earlier accelerator designs optimized for GEMM-dominant machine learning workloads~\cite{tpuv1, eyeriss,dadiannao}.

%% file: sections/3_motivation.tex
\section{The Bimodal Expert Distribution Problem}

\label{sec:motivation}

In this section, we first explain the evolution of modern LLMs, particularly the shift toward increasingly sparse MoE architectures.
Next, we identify how this shift creates a stark disparity in arithmetic intensity across experts and empirically quantify this disparity in recent MoE models.
Finally, we demonstrate why this dynamic invalidates the static scheduling strategies used by existing PIM-enabled systems.

\subsection{Recent LLM Trends}
\label{sec:llmtrends}
The progression from BERT~\cite{bert} to GPT-3~\cite{gpt3} exemplifies how early LLM development pursued greater accuracy primarily through aggressive scaling of dense LLM architectures. However, 
MoE models decouple the growth in total parameter size from per-token computational cost by activating only a small subset of expert parameters for each token~\cite{shazeer2017outrageously, fedus2022switch,du2022glam, lepikhin2020gshard}.
Moreover, the ongoing evolution of MoE models reflects a clear trajectory toward greater sparsity, where models scale to a greater number of experts while activating a smaller fraction per token~\cite{mixtral,meta2025llama4,gptoss,kimiK2}.

Figure~\ref{fig:activated_ratio} visualizes this trend by comparing the activated parameter ratio with the Artificial Analysis Intelligence Index (AAII).
The activated parameter ratio is calculated as the ratio of the activated parameter size to the total parameter size.
AAII is a metric combining multiple dimensions of intelligence where a higher score indicates greater model capability~\cite{artificialanalysis_2025}.
Parameters in non-MoE layers are always activated and are therefore included in the activated parameter size.
We measure the number of activated experts in MoE layers, which depends on input sequences, by running Mixtral-8x22B~\cite{mixtral}, Qwen3-30B-A3B~\cite{qwen3}, Qwen3-Next-80B-A3B~\cite{qwen3}, and GPT-OSS-120B~\cite{gptoss} across various batch sizes ($B$) on traces of real-world requests~\cite{hh-rlhf}.
For simplicity, we hereafter refer to these models as \textit{Mixtral}, \textit{Qwen3}, \textit{Qwen3-Next}, and \textit{GPT-OSS}, and denote the activated parameter ratio as \emph{act-ratio}.

We draw two key observations from Figure~\ref{fig:activated_ratio}.

\begin{tcolorbox}[boxsep=0pt,left=3pt,right=3pt,arc=0pt,colframe=black,colback=mygray,boxrule=0.9pt]
\emph{\textbf{Observation 1:}}
\textit{Modern MoE models with higher capability exhibit lower act-ratios.}
\end{tcolorbox}

Higher-capability MoE models consistently show lower activation ratios across batch sizes, as shown in Figure~\ref{fig:activated_ratio}.
For instance, at batch size $B=1$, the act-ratio is 2.4\% for \textit{GPT-OSS} and 25.0\% for Mixtral-8x22B.
This trend has continued over the past few months: \textit{Qwen3} (released in April 2025) shows higher act-ratios than \textit{Qwen3-Next} (September 2025) and \textit{GPT-OSS} (August 2025).

\begin{tcolorbox}[boxsep=0pt,left=3pt,right=3pt,arc=0pt,colframe=black,colback=mygray,boxrule=0.9pt]
\emph{\textbf{Observation 2:}}
\textit{Modern MoE models exhibit low act-ratios even with larger batch sizes.}
\end{tcolorbox}

Furthermore, even as batch size increases, modern MoE models continue to activate only a small fraction of their parameters, as shown in Figure~\ref{fig:activated_ratio}.
Although the act-ratio of the earlier \textit{Mixtral} rapidly converges to 100\%, that of \textit{GPT-OSS} remains substantially lower.
For example, the median act-ratio for \textit{Mixtral} reaches 100\% at $B=16$, whereas \textit{GPT-OSS} records 47.9\% even at $B=256$.
\textit{Qwen3-Next} also exhibits a similar trend to that of \textit{GPT-OSS}.

In summary, Figure~\ref{fig:activated_ratio} demonstrates that the decreasing act-ratio represents a consistent trend among modern MoE models.
This architectural shift towards greater sparsity fundamentally changes the execution characteristics of MoE serving, invalidating the static workload assumptions that prior PIM systems relied upon.

\subsection{Arithmetic Intensity Disparity}
\label{sec:disparate-ai}

\begin{figure}[thbp]
\begin{center}
\includegraphics[width=\columnwidth]{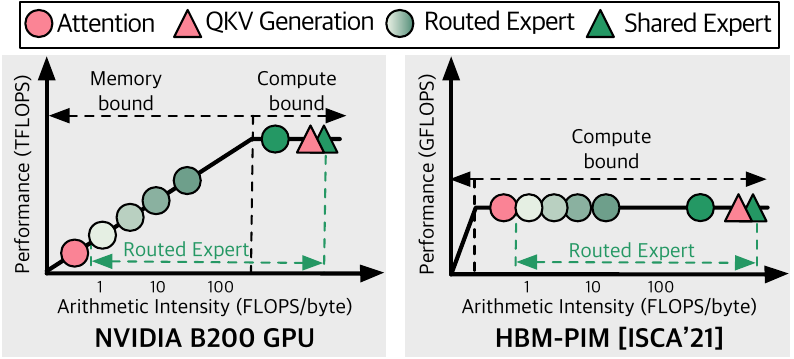}
\Description{Roofline models of the NVIDIA B200 GPU and Samsung HBM-PIM showing arithmetic intensities of operations in MoE models.}
\caption{Roofline models of the NVIDIA B200 GPU~\cite{nvidia-b200-gpu} and the Samsung HBM-PIM~\cite{samsung_hbm2pim}. 
Arithmetic intensities of operations in MoE models are illustrated, where the arithmetic intensity of each routed expert varies with its number of assigned tokens.
Darker colors indicate experts with more routed tokens.
}
\label{fig:roofline}
\end{center}
\end{figure}
\begin{figure*}[htbp]
    \centering
    \includegraphics[width=\textwidth]{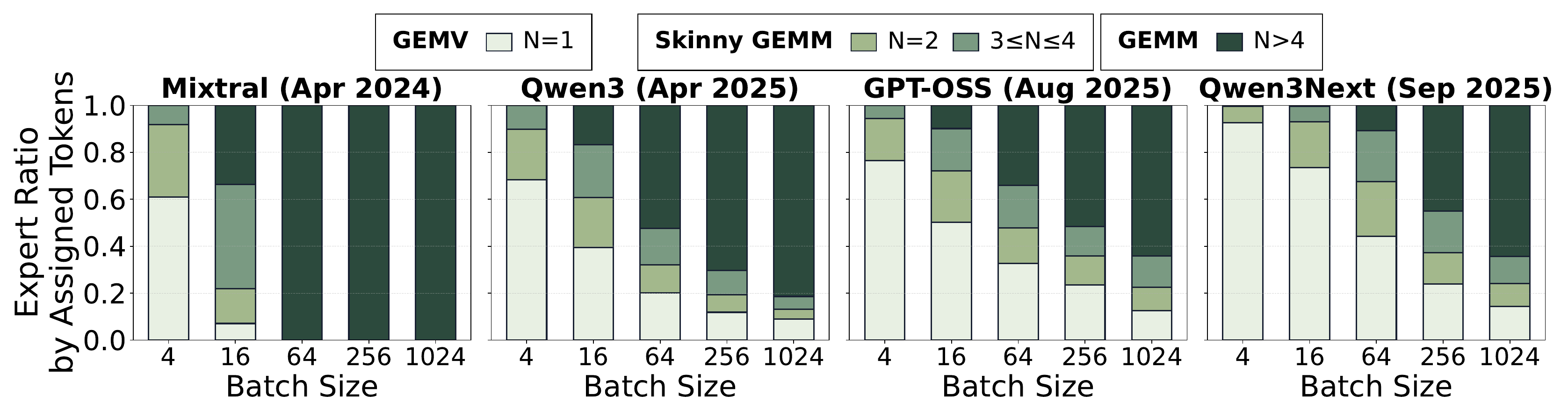}
    \Description{Stacked bar charts showing the proportion of GEMV, skinny GEMM, and GEMM experts across Mixtral, Qwen3, GPT-OSS, and Qwen3-Next at various batch sizes.}
    \vspace{-6mm}
    \caption{Proportion of GEMV, skinny GEMM, and GEMM experts in \textit{Mixtral}, \textit{Qwen3}, \textit{GPT-OSS}, and \textit{Qwen3-Next}, averaged over the HH-RLHF dataset~\cite{hh-rlhf}. 
    $N$ denotes the number of tokens assigned to each expert. 
    For example, in \textit{Qwen3-Next}, the first FFN has a weight matrix of shape $[2048, 512]$, and its computation can be represented as a matrix multiplication of dimensions $[N, 2048]\times[2048, 512]$. 
    The shared expert in \textit{Qwen3-Next} is also included. 
    These bins are used only to expose arithmetic disparity; they are not \moepim scheduling thresholds.
    }
    \label{fig:gemv-ratio-combined}
    
\end{figure*}

As described in Figure~\ref{fig:expert-distribution}, the number of tokens assigned to each expert varies, creating disparities in arithmetic intensity across experts.
To analyze how this disparity affects PIM-enabled systems, Figure~\ref{fig:roofline} presents roofline models of the NVIDIA B200 GPU~\cite{nvidia-b200-gpu} and Samsung HBM-PIM~\cite{samsung_hbm2pim}.
The pink-colored attention and QKV generation operations have fixed arithmetic intensity determined by model configuration and sequence length, so their arithmetic intensities are known before inference begins.
Since the attention operation is typically more memory-bound than other operations on GPUs, offloading the attention operation to PIM can yield substantial performance improvement~\cite{papi,paise,neupims,attacc,duplex}.

In contrast, routed experts exhibit a broad range of arithmetic intensities, as shown by the large variation in arithmetic intensity among the green circles in \Cref{fig:roofline}. 
Experts receiving only a few tokens have low arithmetic intensity and are inefficiently executed on GPU.
On the other hand, experts with many tokens are compute-bound and achieve higher efficiency on GPU than on PIM.
Consequently, assigning experts to PIM without considering their arithmetic intensity
can lead to severe inefficiency.

To fully exploit PIM for efficient MoE serving, these disparities must be carefully addressed. 
Figure~\ref{fig:roofline} provides a straightforward insight: experts with low arithmetic intensity are best executed on PIM, whereas those with high intensity should be processed on GPU.
In the next subsection, we quantify the degree of disparity in arithmetic intensity across experts to analyze its performance implications for PIM-enabled systems.

\subsection{Quantifying Arithmetic Disparity}
\label{sec:quantify-disparity}

We draw two key observations from Figure~\ref{fig:gemv-ratio-combined}, which illustrates the degree of disparity in arithmetic intensity across experts. 

\begin{tcolorbox}[boxsep=0pt,left=3pt,right=3pt,arc=0pt,colframe=black,colback=mygray,boxrule=0.9pt]
\emph{\textbf{Observation 3:}}
\textit{A substantial fraction of expert computations remain memory-bound even with large batch sizes.}
\end{tcolorbox}

The number ($N$) of tokens assigned to each expert determines its arithmetic intensity because all experts within a MoE layer share identical parameter tensor dimensions~\cite{gemini2.5,qwen3,gptoss,kimiK2,deepseekR1,deepseekV3,glm4.5}. 
Accordingly, we classify experts into two groups:
(1) memory-bound (unpopular) experts performing GEMV or skinny GEMM operations, and
(2) compute-bound (popular) experts performing GEMM operations. We bin experts into $N=1$, $N=2$, $3\le N\le 4$, and $N>4$ to characterize arithmetic intensity in Figure~\ref{fig:gemv-ratio-combined}.
For shared experts in MoE models such as \textit{Qwen3-Next}~\cite{qwen3}, $N$ equals the batch size ($B$), typically resulting in compute-bound operations when $B > 4$.

As expected, memory-bound expert computations dominate at small batch sizes.
For example, when $B=4$, 92.5\% of expert computations in \textit{Qwen3-Next} correspond to GEMV.
Moreover, a significant portion of expert computations in modern MoE models remains memory-bound even at larger batch sizes.
The earlier MoE model \textit{Mixtral} shows almost no memory-bound experts once $B\ge64$.
On the other hand, as we move to newer MoE models, the ratio of memory-bound expert computations grows notably. 
Newer-generation MoE models show a drastic difference: even at $B=64$, 47.6\% of expert computations in \textit{Qwen3}, 89.3\% in \textit{Qwen3-Next} and 65.9\% in \textit{GPT-OSS} are memory-bound.
This high ratio persists even when $B=256$, where 50.1\% and 56.6\% of experts in the respective models remain memory-bound.
Moreover, a striking finding from Figure~\ref{fig:gemv-ratio-combined} is that a large portion of expert computations continues to be memory-bound even in large-batch scenarios for recent models, which are designed to improve throughput. 

\begin{tcolorbox}[boxsep=0pt,left=3pt,right=3pt,arc=0pt,colframe=black,colback=mygray,boxrule=0.9pt]
\emph{\textbf{Observation 4:}}
\textit{Single-token assignments (GEMV) occur frequently, indicating a high proportion of low arithmetic intensity computations.}
\end{tcolorbox}

The difference in single-token assignments across generations of MoE models is particularly noteworthy, as these computations degenerate into GEMV operations. 
We refer to experts that are assigned exactly one token in a batch as \emph{GEMV experts}, since computation in the experts is composed of GEMV operations.
At $B=64$, GEMV experts account for 20.2\% of expert computations in \textit{Qwen3}, 32.6\% in \textit{GPT-OSS}, and 44.2\% in \textit{Qwen3-Next}.
This trend persists even at $B=256$, where the corresponding ratios are 11.9\%, 23.5\%, and 23.9\%, respectively.
Although Figure~\ref{fig:gemv-ratio-combined} extends to $B=1024$, batch sizes below 256 are generally preferred to avoid excessive latency and potential SLO violations~\cite{attacc,neupims,duplex,papi,paise,specpim}.

In summary, the large disparity in arithmetic intensity across experts calls for an adaptive acceleration approach, which our framework is designed to provide.

\subsection{Limitations of Prior PIM-enabled Systems}
\label{sec:whatismissing}

\begin{figure}
    \centering
    \includegraphics[width=\columnwidth]{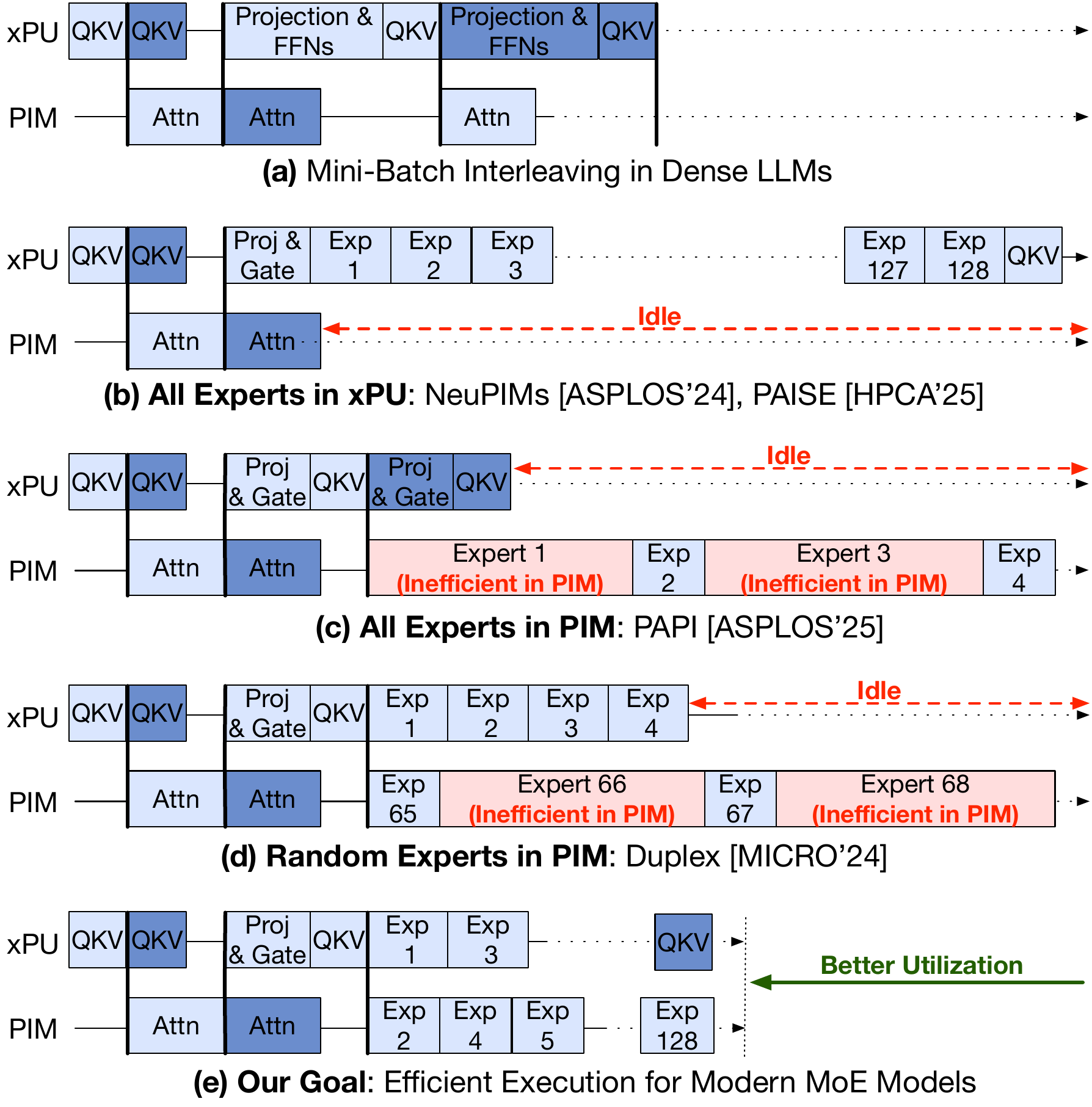}
    \Description{Execution flow diagrams for PIM-enabled systems showing dense LLM and MoE model execution patterns.}
    \caption{Execution flow examples in PIM-enabled systems for dense LLMs and MoE models. We use xPU as a generalized term for the host processor because prior work targets different compute substrates, including GPUs and NPUs.
    }
    \label{fig:motiv-executionflow-colored}
\end{figure}

Prior PIM-enabled systems mainly target memory-bound operations in dense LLMs~\cite{papi,neupims,attacc,duplex,paise,pimisallyouneed}.
As discussed in Section~\ref{sec:pim-for-llms}, these systems commonly offload attention operations during the decoding phases to PIM, because attention has low arithmetic intensity and benefits from the high internal bandwidth of PIM.
In contrast, compute-intensive operations such as FFNs and QKV generation remain on the GPU.
To overlap computations between GPU and PIM, most prior work splits each batch ($N$) into two mini-batches ($N/2$) and interleaves their execution across the two devices, as illustrated in Figure~\ref{fig:motiv-executionflow-colored} (a). 
This interleaving reduces the idle time of both PIM and GPU when running dense LLMs.
A naive sequential execution for dense LLMs in systems composed of GPUs and PIM would offload the attention operation to PIM.
The mini-batch interleaving strategy utilizes the fact that dependencies exist only within a mini-batch, and it improves the throughput of PIM-enabled systems by reducing idle times.
Furthermore, since serialized execution limits throughput by leaving one device idle, prior work interleaves two mini-batches, executing one on the GPU and the other on the PIM concurrently, as described in Figure~\ref{fig:motiv-executionflow-colored} (a)~\cite{duplex,neupims}.

However, these strategies become inefficient when applied to modern MoE models. 
As illustrated in Figure~\ref{fig:motiv-executionflow-colored} (b), PIM remains idle during expert computation in systems that offload only attention operations to PIM~\cite{neupims,paise}.
Although GPUs can process multiple experts using batched matrix multiplication instead of executing each expert sequentially, 
the overhead of loading all activated expert parameters from off-chip memory remains substantial, preventing full PIM utilization.
Similarly, as shown in Figure~\ref{fig:motiv-executionflow-colored} (c), offloading all expert computations to PIM causes the GPU to remain idle~\cite{papi}.
Therefore, to fully exploit PIM-enabled systems, the expert layer should be co-processed across both PIM and GPU.

As described in Figure~\ref{fig:motiv-executionflow-colored} (d), naively co-processing FFNs on both PIM and GPU also introduces inefficiency in modern MoE models.
Prior work evaluates such co-processing by assuming a uniform expert distribution~\cite{duplex}, which was valid for earlier MoE models such as \textit{Mixtral} with large batch sizes.
In contrast, as elaborated in Section~\ref{sec:quantify-disparity}, modern MoE models exhibit highly imbalanced token-to-expert assignments even with large batches. 
Executing experts with many assigned tokens on PIM becomes inefficient because commercial PIM architectures are optimized for GEMV rather than GEMM, resulting in lower throughput.

In summary, existing PIM strategies designed for dense LLMs and earlier MoE models are ineffective for modern MoE models.
Figure~\ref{fig:motiv-executionflow-colored} (e) illustrates the execution flow that this work aims to achieve, where MoE serving is optimized to better utilize both PIM and GPU resources for higher throughput.

%% file: sections/4_overview.tex
\section{Overview}
\label{sec:overview}

\begin{figure}[t]
    \centering
    \includegraphics[width=0.995\columnwidth]{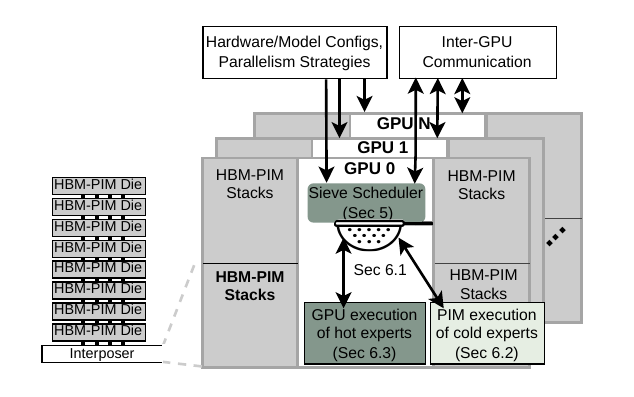}
    \Description{Overview diagram of the Sieve framework showing hardware configuration and runtime scheduling of GPU and PIM computations.}
    \caption{An overview of \moepim. Hardware and model configurations are determined before serving MoE models. The \moepim scheduler leverages the runtime-determined distribution of tokens across experts to enable efficient co-execution of GPU and PIM computations.
    }
    \label{fig:overview-sieve}
\end{figure}

Building on the aforementioned insights, we introduce \moepim, a dynamic expert-aware PIM acceleration framework for serving MoE models.
\moepim leverages the disparity in arithmetic intensity across experts and the complementary characteristics of GPUs and PIM.
\moepim requires no hardware modifications: no new PIM commands and no changes to the existing PIM architecture or command interface. 
Therefore, \moepim incurs no additional hardware cost, such as area overhead. 
Instead, \moepim improves the utilization of existing PIM resources by exploiting the interaction between PIM architectural characteristics and modern MoE serving workloads.

\textbf{Memory Model: } 
A PIM die is a conventional HBM die augmented with PIM processing units, trading a portion of memory capacity for compute capability while maintaining approximately the same die area as a standard HBM die. 
Expert parameters are stored in off-chip HBM-PIM dies across GPUs, depending on the expert-parallelism and tensor-parallelism configuration. 
Expert migration across GPUs may arise when expert-parallel load balancing (EPLB) is enabled, which can be utilized with \moepim. 
For GPU execution, expert parameters are fetched from off-chip memory into GPU on-chip memory before computation. 
For PIM execution, PIM processing units directly access expert parameters from local DRAM banks, without transferring them to GPU on-chip memory. 
Accordingly, \moepim determines \emph{that each expert is executed on either PIM or GPU}, rather than where its parameters are stored.

\textbf{Interconnect Model: } 
Prior work assumes a global interconnect across multiple GPUs (or NPUs) and PIM devices~\cite{neupims, attacc, papi, pimoe}.
Under this assumption, any GPU can access any PIM device.
This abstraction fails to capture inter-GPU communication overhead caused by expert parallelism across GPUs, even though such overhead is a key bottleneck in MoE inference.

We instead assume that each GPU integrates multiple HBM-PIM stacks, where a stack consists of multiple HBM-PIM dies as described in Figure~\ref{fig:overview-sieve}.
Each HBM-PIM die is accessible only through its attached GPU.
Therefore, in this setting, a token cannot access an expert in a remote HBM-PIM stack without going through the corresponding remote GPU.
This design matches real-world multi-GPU systems, where GPUs are interconnected via NVLink but HBM dies are not directly connected to other GPUs.

\textbf{\moepim System: } 
Executing all experts on GPUs incurs significant data movement overhead for transferring all activated experts' parameters from HBM to GPU before computation.
In contrast, executing all experts on PIM eliminates this transfer cost but reduces overall performance due to PIM's lower computational throughput compared to GPUs.
A static assignment of experts to PIM is also inefficient, as the number of tokens assigned to each expert varies dynamically across batches at runtime.
Therefore, efficient execution of modern MoE models requires adaptively partitioning expert workload between GPUs and PIM at runtime.

\moepim adopts a dynamic approach that partitions experts into two groups based on their arithmetic intensity.
Memory-bound experts with low arithmetic intensity are executed on PIM, while compute-bound experts with high intensity are executed on GPUs.
By executing certain experts on PIM, \moepim reduces off-chip memory access overhead.
Moreover, \moepim dynamically decides the partition while accounting for the overhead of inter-GPU communication and the execution of attention operations on PIM.

Figure~\ref{fig:overview-sieve} shows an overview of \moepim.
Based on the runtime token-to-expert distribution and the overhead of inter-GPU communication and PIM-based attention operations, the Sieve scheduler (Section~\ref{sec:sieving}) efficiently partitions expert execution between GPUs and PIM. Section~\ref{sec:sieve-system} then presents the \moepim system, which realizes this partitioning through three key components: lightweight coordination across GPUs and their attached HBM-PIM stacks (Section~\ref{sec:synchronization}), efficient execution of unpopular experts on PIM (Section~\ref{sec:pimfriendly}), and efficient execution of popular experts on GPUs (Section~\ref{sec:xpufriendly}).

%% file: sections/5_scheduler.tex
\section{\moepim Scheduler}
\label{sec:sieving}
The \moepim scheduler is a runtime scheduler for MoE models.
It dynamically partitions expert computations between GPUs and their attached HBM-PIM stacks.
This subsection describes how the \moepim scheduler determines an effective scheduling strategy, accounting for factors such as the MoE architecture, hardware configuration, and parallelization strategies, including data, context, expert, and tensor parallelism.

\subsection{Objective Function}
The \moepim scheduler leverages the arithmetic intensity disparity caused by the bimodal token--expert distribution at runtime to make efficient scheduling decisions.
To maximize throughput and minimize latency, the \moepim scheduler follows three core principles:
\circnum{1} It jointly considers hardware utilization across four resources: the interconnect, PIM, GPU compute, and GPU off-chip memory bandwidth.
\circnum{2} Popular experts are executed on GPUs to avoid performance degradation.
\circnum{3} Unpopular experts with low arithmetic intensity are offloaded to PIM.
Its decisions are guided by the runtime expert distribution, hardware configuration, and MoE model configuration.

Principle \circnum{1} defines the primary objective of the \moepim scheduler, while \circnum{2} and \circnum{3} specify how the \moepim scheduler achieves it.
We formalize principle \circnum{1} with the objective function in Equation~\ref{eq:objective-function}. 
Let $E$ denote the set of activated experts on a GPU and its corresponding HBM-PIM stacks, $S \subseteq E$ the subset assigned to PIM, and $G = E - S$ the subset assigned to the GPU.
The \moepim scheduler finds the partition $S^{*}$ that minimizes the bottleneck across three components:

\begin{equation}\label{eq:objective-function}
S^{*} = \arg\min_{S \subseteq E} \, \max\!\Big(T_{\mathrm{Comm}},\; T_{\mathrm{GPU}}(G),\; T_{\mathrm{PIM}}(S)\Big)
\end{equation}

Prior work has proposed highly fine-grained overlap of computation and communication to maximize GPU utilization during MoE inference~\cite{comet,megascale-infer}.
Accordingly, the \moepim scheduler uses the maximum estimated execution time across the interconnect, GPU, and PIM.
Since the \moepim scheduler runs on the critical path, it is designed to identify the dominant bottleneck with low overhead rather than exactly predict execution time. The detailed execution times reported in Section~\ref{sec:main-result} are obtained from cycle-accurate simulation.

$T_{\mathrm{Comm}}$ denotes the estimated inter-GPU communication time incurred by tensor and expert parallelism. 
Because tokens are routed to other GPUs totally based on the gating results regardless of the \moepim scheduler's partitioning decision, $T_{\mathrm{Comm}}$ is independent of both $S$ and $G$.
$T_{\mathrm{GPU}}(G) = \max\big(T_{\mathrm{offchip}}(G), T_{\mathrm{comp}}(G)\big)$ denotes the estimated execution time for GPU operations, such as popular experts retained on the GPU by principle \circnum{2}. 
The \moepim scheduler determines it as the larger of $T_{\mathrm{offchip}}(G)$ (the off-chip memory access time for loading parameters and storing intermediate values) and $T_{\mathrm{comp}}(G)$ (the GPU computation time).
$T_{\mathrm{PIM}}(S)$ denotes the estimated execution time for PIM operations, such as attention and unpopular experts assigned to PIM by principle \circnum{3}.

\textbf{Timing Models:}
The following estimates for $T_{\mathrm{GPU}}(G)$, $T_{\mathrm{PIM}}(S)$, and $T_{\mathrm{Comm}}$ are used solely within the \moepim scheduler to guide the partitioning decision, while the execution times reported in Section~\ref{sec:main-result} are obtained from cycle-accurate simulation.
Since the Sieving algorithm runs on the critical path, we prioritize lightweight estimates over precise modeling to keep its overhead low.
 
$T_{\mathrm{offchip}}(G)$ is computed by dividing the amount of data transferred between HBM-PIM and the GPU, including parameters and intermediate values, by the HBM bandwidth.
$T_{\mathrm{comp}}(G)$ is estimated by dividing the number of GPU operations, including all operations except decode-phase attention and PIM-side expert computation ($S$), by the GPU's peak compute throughput.
Specifically, the \moepim scheduler assumes that GPU-side expert computation is efficiently executed using grouped GEMM~\cite{vllm, sglang}, where the experts in $G$ are batched into a single kernel with variable group sizes determined by token counts.
$T_{\mathrm{Comm}}$ is estimated by dividing the amount of inter-GPU communication incurred by tensor and expert parallelism by the interconnect bandwidth.

For $T_{\mathrm{PIM}}(S)$, we follow the assumption in prior work that PIM executes a multi-token expert as serialized GEMV operations through the DRAM command interface~\cite{neupims, samsung_hbm2pim}.
However, prior work has shown that these overheads are non-linear~\cite{paise}.
Executing a 1-token expert on PIM does not take half the time of executing a 2-token expert, mainly because of DRAM timing overheads such as row buffer conflicts, bank contention, and refresh cycles.
As a result, a roofline-based estimate can overestimate expert execution time on PIM by 1.8--4.2$\times$, because it does not capture DRAM timing overhead.

To address this, the \moepim scheduler maintains a runtime cost table whose keys are token counts and whose values are the observed PIM execution times for experts with those token counts.
After each iteration, the \moepim scheduler updates the cost table using an exponential moving average of the observed PIM GEMV execution times.
For token counts that have not yet been observed, the \moepim scheduler uses a roofline estimate obtained by dividing the number of operations by the PIM's peak compute throughput.
Although this estimate may be inaccurate, the scheduler uses it only as a one-time fallback until an observed PIM timing estimate becomes available.
The PIM cost table converges within the first few iterations, as the varying expert distributions across successive batches quickly populate entries for the relevant token counts.

\subsection{\moepim Scheduling Algorithm}
As defined in Equation~\ref{eq:objective-function}, the \moepim scheduler seeks to identify an optimal $S^{*} \subseteq E$ such that the estimated execution time is minimized.
However, an exhaustive search across all $2^{|E|}$ combinations is computationally infeasible; for instance, Qwen3-Next-80B-A3B would require evaluating $2^{512}$ combinations.

To find $S^{*}$ efficiently, the \moepim scheduler employs a greedy heuristic. 
First, it sorts all experts in descending order based on their token counts.
The \moepim scheduler assumes all expert computations are initially assigned to PIM.
It then iteratively evaluates whether moving the computation of the expert with the highest token count from PIM to the GPU would reduce the estimated overall execution time  $T_{\mathrm{total}}=\max\!\Big(T_{\mathrm{Comm}},\; T_{\mathrm{GPU}}(G),\; T_{\mathrm{PIM}}(S)\Big)$.
If $T_{\mathrm{total}}$ decreases, the expert computation is reassigned to the GPU.
Assigning the most popular expert to the GPU yields the largest reduction in $T_{\mathrm{PIM}}$, while the GPU cost of executing an additional expert is relatively constant as most experts are memory-bound on the GPU.
The \moepim scheduler continues until moving the next expert increases $T_{\mathrm{total}}$, indicating that the remaining experts have sufficiently low arithmetic intensity to be better handled by PIM.
This process also stops if all expert computations have been assigned to the GPU, which is a common outcome in MoE models with a small number of experts such as Mixtral-8x7B.

\textbf{Overhead of the \moepim Scheduler: }
The \moepim scheduler is designed to be lightweight and practical for modern MoE models where $|E| < 1024$.
By prioritizing the most popular experts, the scheduler achieves the most significant reductions in $T_{\mathrm{PIM}}$ with minimal iterations.
Since the complexity is dominated by a single sort and a linear scan over the expert list, the computational overhead of the \moepim scheduler is negligible.
For example, the \moepim scheduler runs in approximately $20\mu\mathrm{s}$ on a B200 GPU for a MoE layer of DeepSeek-R1, even without kernel optimization.
In our evaluation, each GPU executes this algorithm locally after receiving global token counts via the AllGather step explained in \Cref{sec:synchronization}, and this overhead is fully accounted for in our results.

\textbf{Comparison with PIMoE~\cite{pimoe}: }
PIMoE uses a static \textit{threshold} to assign popular experts with $N \geq \textit{threshold}$ to the NPU when the estimated NPU execution time is lower than the estimated PIM execution time.
However, PIMoE partitions expert computations without accounting for attention operations that already occupy PIM. 
As attention time on PIM grows with longer sequences or higher request concurrency, a partition that appears balanced for an MoE operation can become inefficient in end-to-end MoE inference. This is because the PIM execution time increases substantially, turning PIM into the bottleneck.

PIMoE also assumes a global interconnect that allows an NPU to access any PIM device with uniform latency. 
This assumption ignores a key overhead in real-world multi-GPU systems: the inter-GPU communication required for token dispatch and combination in expert parallelism.
When PIMoE is adapted to multi-GPU systems in which each GPU integrates multiple HBM-PIM stacks, the PIM execution cost exceeds both the inter-GPU communication cost and the GPU execution cost.
In this case, assigning more experts to the GPU can reduce PIM execution time and increase GPU utilization.
Therefore, rather than relying on a static threshold, the \moepim scheduler must dynamically consider both inter-GPU communication and attention time on PIM.

%% file: sections/6_system.tex
\section{\moepim System}
\label{sec:sieve-system}

\begin{figure*}[t]
    \centering
    \includegraphics[width=\textwidth]{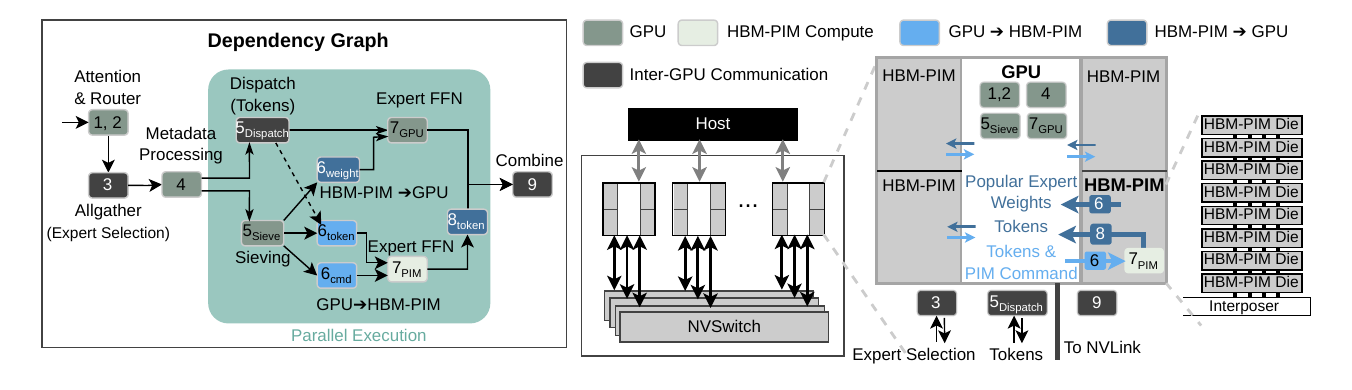}
    \Description{Overview of the Sieve system showing dependencies across operations between GPUs and HBM-PIM stacks.}
    \caption{An overview of the \moepim System and the dependencies across operations.
    }
    \label{fig:sieve-execution-flow}
\end{figure*}

This section describes the \moepim runtime system, which turns the \moepim scheduler's per-expert placement decisions into an executable MoE-layer pipeline. The runtime must coordinate routing metadata, token dispatch, GPU/PIM execution, and final aggregation across GPUs and their attached HBM-PIM stacks. We first describe this cross-device coordination in \Cref{sec:synchronization}, then discuss the PIM and GPU execution paths in \Cref{sec:pimfriendly} and \Cref{sec:xpufriendly}.

\subsection{Coordinating GPUs and HBM-PIM}
\label{sec:synchronization}

\moepim coordinates each MoE layer as a dependency graph over routing, communication, and expert execution. 
This graph exposes independent operations that can be overlapped across inter-GPU communication, GPU compute, and HBM-PIM compute, while preserving the ordering required for correctness.

\textbf{Dependencies across operations:}
An example dependency graph for co-executing MoE layers on a PIM-enabled multi-GPU system is shown in the left panel of \Cref{fig:sieve-execution-flow}.
From the attention output (\blackcircle{1}), the router (\blackcircle{2}) computes a token-to-expert routing map.
With data-parallel attention, each GPU initially holds only a local routing map containing a disjoint subset of tokens.
Since expert parallelism is used across GPUs, an AllGather step (\blackcircle{3}) aggregates these local maps into a global view. This allows each GPU to identify where the tokens assigned to each expert reside.
During a metadata processing step (\blackcircle{4}), each GPU prepares fixed-size tensors per expert, enabling efficient grouped GEMM execution on the GPU.

After this stage, \moepim enables parallel co-execution on GPUs and PIM to efficiently utilize multiple resources: inter-GPU bandwidth, GPU compute, PIM compute, and HBM-PIM bandwidth. 
First, tokens are dispatched across GPUs (\blackcircle{5}$_{\mathrm{Dispatch}}$) so that each token resides on the GPU hosting its assigned expert. In parallel, the \moepim scheduler explained in \Cref{sec:sieving} runs on each GPU (\blackcircle{5}$_{\mathrm{Sieve}}$) to determine which experts should execute on the GPU versus PIM.

If the \moepim scheduler assigns an expert to the GPU, its parameters are loaded from HBM-PIM to GPU (\blackcircle{6}$_{\mathrm{HBM\text{-}PIM\rightarrow GPU}}$). 
Since shared experts~\cite{gptoss, qwen3, deepseekmoe, deepseekR1} receive every token and thus lead to a large GEMM, we start loading the weights for these shared experts right after \blackcircle{4}. 
This enables more overlap by relaxing the dependency \blackcircle{4}$\rightarrow$\blackcircle{5}$_{\mathrm{Dispatch}}$ $\rightarrow$\blackcircle{6}$_{\mathrm{weight}}$ for shared experts.
As an optimization, parameter loading can also begin early for experts assigned multiple tokens within the local routing map.
Once loaded, these experts are processed via grouped GEMM on the GPU (\blackcircle{7}$_{\mathrm{GPU}}$).
If \moepim assigns an expert to PIM, the GPU issues PIM commands to HBM-PIM while sending the tokens
(\blackcircle{6}$_{\mathrm{GPU\rightarrow HBM\text{-}PIM}}$).
The expert FFN is then executed directly on PIM (\blackcircle{7}$_{\mathrm{HBM\text{-}PIM}}$) and the produced tokens get loaded back to the GPU (\blackcircle{8}).

To summarize, the popular experts are executed on the GPU through \blackcircle{1}$\rightarrow$\blackcircle{2}$\rightarrow$\blackcircle{3}$\rightarrow$\blackcircle{4}$\rightarrow$\blackcircle{5}$_\mathrm{Sieve}$$\rightarrow$\blackcircle{6}$_{\mathrm{HBM\text{-}PIM\rightarrow GPU}}$$\rightarrow$\blackcircle{7}$_{\mathrm{GPU}}$$\rightarrow$ \blackcircle{9} with an additional dependency (\blackcircle{5}$_\mathrm{Dispatch}$$\rightarrow$\blackcircle{7}$_{\mathrm{GPU}}$) for tokens dispatched from remote GPUs.
The unpopular experts are executed on PIM through \blackcircle{1}$\rightarrow$\blackcircle{2}$\rightarrow$\blackcircle{3}$\rightarrow$\blackcircle{4}$\rightarrow$\blackcircle{5}$_\mathrm{Sieve}$$\rightarrow$\blackcircle{6}$_{\mathrm{GPU\rightarrow HBM\text{-}PIM}}$$\rightarrow$\blackcircle{7}$_{\mathrm{HBM\text{-}PIM}}$ $\rightarrow$\blackcircle{8}$\rightarrow$\blackcircle{9} with an additional dependency (\blackcircle{5}$_\mathrm{Dispatch}$$\rightarrow$\blackcircle{6}$_{\mathrm{GPU\rightarrow HBM\text{-}PIM}}$) for tokens dispatched from remote GPUs.

\textbf{Synchronization: }
Since \moepim executes unpopular experts by converting each of them into a GEMV operation, the number of GEMV operations on PIM can vary across batches.
However, the dimensionality of each GEMV remains constant for a given MoE model.
This property allows the memory controller to schedule PIM commands deterministically~\cite{neupims}.
As a result, the sequence \blackcircle{6}$_{\mathrm{GPU\rightarrow HBM\text{-}PIM}}$→\blackcircle{7}$_{\mathrm{HBM\text{-}PIM}}$→\blackcircle{8} proceeds without violating DRAM timing parameters such as refresh intervals.
Moreover, to ensure that \blackcircle{9} starts only after both \blackcircle{7}$_{\mathrm{HBM\text{-}PIM}}$ and \blackcircle{7}$_{\mathrm{GPU}}$ have completed, \moepim encodes these data dependencies within its DAG representation, following prior work~\cite{neupims}.

\textbf{Aggregation: }
Depending on the parallelization strategy in a system, additional aggregation between GPU and HBM-PIM may be required if some tokens dispatched to a GPU are computed on its corresponding HBM-PIM.
Each dispatched token and the result of its expert computation have dedicated on-chip memory addresses after \blackcircle{5}$_\mathrm{Dispatch}$.
Therefore, although the token's expert computation is conducted in PIM and its result is loaded back to the GPU (\blackcircle{6}$_{\mathrm{GPU\rightarrow HBM\text{-}PIM}}$$\rightarrow$\blackcircle{7}$_{\mathrm{HBM\text{-}PIM}}$$\rightarrow$\blackcircle{8}), the result can be stored again at the dedicated on-chip memory address.
After that, \blackcircle{9} is performed to send the dispatched token to the original GPU, where the aggregation of the token's intermediate values from multiple expert computations is executed.

\subsection{Executing Unpopular Experts on PIM}
\label{sec:pimfriendly}

After the \moepim scheduler decides the expert computation orchestration, PIM performs computation for GEMV and skinny GEMM experts.
\moepim introduces a PIM-friendly MoE execution model for these experts.

\textbf{Issuing PIM Commands: }
To support heterogeneous execution between GPUs and PIM, 
\moepim employs a custom GPU kernel that initializes and controls PIM operations.
This kernel issues PIM commands using tensor sizes and memory addresses determined dynamically at runtime.
As a result, \moepim can be realized entirely in software on GPUs. 
This design is feasible because prior studies have extended GPU programming models to allow custom kernels to manage PIM operations~\cite{pimba, neupims, duplex}.

As shown in \Cref{fig:sieve-execution-flow}, expert computation on PIM involves three sub-steps:
(i)~distributing the token tensor to the global buffers of all PIM channels and activating the corresponding rows of the operand matrix in the row buffers for PIM computation (\blackcircle{6}$_{\mathrm{token}}$),
(ii)~issuing the GEMV operation (\blackcircle{6}$_{\mathrm{cmd}}$) to perform expert FFN computation via a series of dot products (\blackcircle{7}$_{\mathrm{HBM\text{-}PIM\rightarrow GPU}}$), and
(iii)~reading the GEMV results back from PIM to GPU on-chip memory (\blackcircle{8}$_{\mathrm{token}}$).
\moepim can be implemented on any PIM architecture or command interface that supports these three sub-steps.
For example, in NeuPIMs~\cite{neupims}, sub-step~(i) is implemented with the \textit{PIM\_GWRITE} command, and sub-steps~(ii) and~(iii) are implemented with the \textit{PIM\_GEMV} command.

When the custom GPU kernel triggers PIM execution, it issues PIM commands whose arguments are computed based on the output of the \moepim scheduler.
Since unpopular experts are executed through separate PIM commands, the arguments differ across GEMV operations.
For example, the arguments include the row and column indices of the memory array that stores expert parameters for GEMV computation in sub-step~(ii), and the GPU on-chip memory address used to load the results back in sub-step~(iii).
These addresses vary across PIM channels because each channel produces results for a different portion of the expert output.
Preparing these arguments requires only basic arithmetic operations, which can be performed on the GPU at runtime, as discussed in prior work~\cite{pimba, neupims, duplex}.

\textbf{Converting Skinny GEMM to GEMV Operations: }
Commercial PIM designs are optimized for dot-product operations rather than GEMM operations~\cite{hynix_pim_isscc22, newton, samsungPIM_hcs23}.
To align with these architectures, \moepim converts each skinny GEMM expert into a sequence of equivalent GEMV operations and issues the corresponding PIM commands.
For example, if an expert is selected by three tokens, \moepim performs three GEMV operations on PIM.
In this case, sub-steps~(i), (ii), and~(iii) are executed sequentially on PIM for each token associated with that expert.

\textbf{Parallelizing Expert Computation on PIM: } 
\moepim adopts tensor parallelism across PIM channels to maintain high utilization even when the expert distribution is highly imbalanced.
In other words, each expert’s parameters are evenly sharded across all PIM channels, allowing every GEMV operation to be divided across the channels.
Moreover, the parameters of a given expert are aligned to identical indices across banks and channels, which reduces the address calculation overhead.
As a result, all PIM resources can be efficiently utilized even under dynamically changing and imbalanced expert distributions.
Since this form of parallelism uses existing PIM commands, it remains compatible with current DRAM-based PIM interfaces used in prior work~\cite{neupims, attacc}.

An alternative is expert parallelism (EP), which executes different experts concurrently on separate hardware components because expert computations are independent~\cite{vllm, sglang, ktransformers}. 
In PIM architectures, EP can be applied at multiple granularities, such as banks, bank groups, and channels.
However, assigning individual banks or bank groups to distinct experts is inefficient because all processing units within a PIM channel must share the same vector operand during each GEMV operation~\cite{newton, samsung_hbm2pim, neupims,paise}.
Banks or bank groups without activated experts for a given token remain idle, leading to low utilization.
Although channel-level EP can leverage inter-bank parallelism, it also risks low utilization when some channels contain experts selected by very few tokens, as discussed in Section~\ref{sec:disparate-ai}. 
Supporting cross-channel or cross-bank access would require dedicated hardware mechanisms that are rarely available in commercial PIM systems~\cite{psyncpim}. 
Therefore, \moepim does not adopt EP because it cannot ensure high and balanced PIM utilization.

Prior work has also explored distributing the attention operation and KV cache of each request across PIM channels~\cite{neupims, duplex}.
However, distributing expert computation in the same manner is inefficient.
Tokens from different requests may require the same expert parameters, which demands cross-channel access or redundant copies of expert parameters in multiple channels.
Consequently, offloading expert computations to PIM requires a parallelization strategy that differs from attention.

\subsection{Executing Popular Experts on GPUs}
\label{sec:xpufriendly}

After running the \moepim scheduler described in Section~\ref{sec:sieving}, \moepim identifies the popular experts and their assigned tokens.
To execute these GEMM experts on GPUs and store their intermediate results in GPU on-chip memory, \moepim follows the common practice in state-of-the-art LLM serving frameworks~\cite{vllm, sglang, ktransformers, huggingfacetransformers}.
At \blackcircle{7}$_\mathrm{GPU}$ in \Cref{fig:sieve-execution-flow}, \moepim performs the computation for popular experts on GPUs using grouped GEMM or batch matrix multiplication.
The outputs of all popular experts are written into a contiguous region of GPU on-chip memory.
After the outputs of all unpopular experts have also been transferred from PIM to GPU on-chip memory, \moepim reorders the expert-grouped results into token-grouped results.
This permutation enables the final aggregation that computes the weighted sum for each token, yielding the MoE layer output.

%% file: sections/7_evaluation.tex
\section{Evaluation}
\label{sec:eval}
\subsection{Methodology}
\label{sec:methodology}
\input{tab_memconfig_b200gpu}

\begin{figure*}[t]
    \centering
    \includegraphics[width=\textwidth]{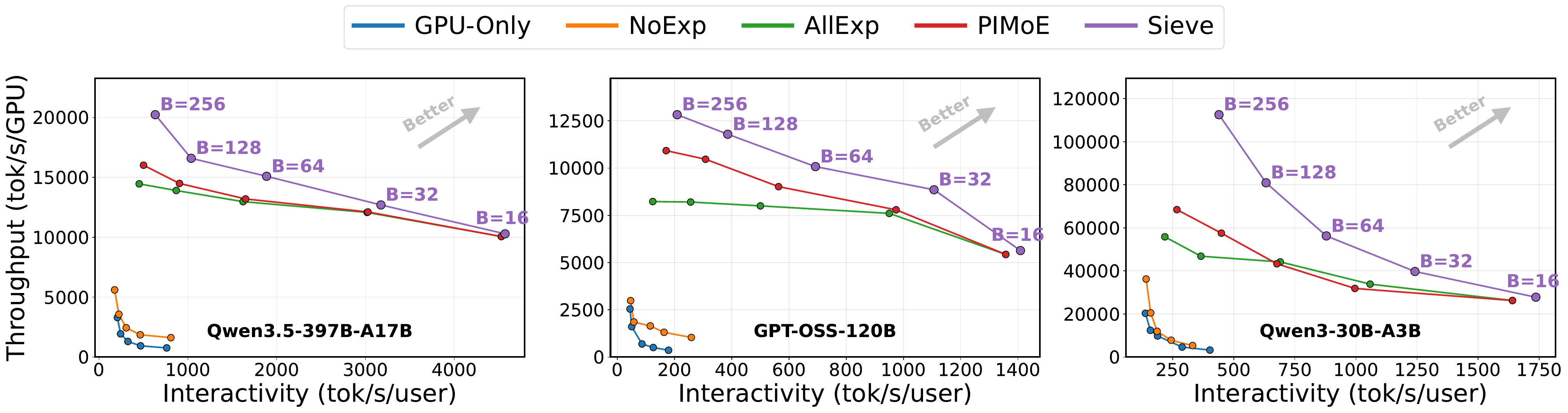}
    \Description{Pareto curves comparing throughput and interactivity of GPU-Only, NoExp, AllExp, PIMoE, and Sieve across GPT-OSS, Qwen3.5, and Qwen3.}
    \caption{Evaluation of throughput and interactivity achieved by \textit{GPU-Only}, \textit{NoExp}~\cite{neupims,paise}, \textit{AllExp}~\cite{papi,stratum}, \textit{PIMoE}~\cite{pimoe}, and \moepim.
    \textit{NoExp} and \textit{AllExp} execute all expert computation on GPUs and PIM, respectively.
    \textit{PIMoE} uses a static threshold to assign expert computation to GPUs and their attached HBM-PIM stacks.
    Four B200 GPUs, eight B200 GPUs, and one B200 GPU are used for \textit{GPT-OSS}, \textit{Qwen3.5}, and \textit{Qwen3}, respectively, where each GPU has its own HBM-PIM stacks.}
    \label{fig:main-result}
\end{figure*}

\textbf{Simulation Methodology: } 
We develop a cycle-accurate simulator using Ramulator 2.0~\cite{ramulator2} and Duplex~\cite{duplex} to evaluate the performance of \moepim.
Table~\ref{tab:memconfig_b200} summarizes the GPU and HBM-PIM configurations used for detailed simulations of DRAM and PIM commands, following prior work~\cite{pimba, neupims, duplex}.
Our PIM implementation includes state-of-the-art PIM hardware components for LLMs, such as dual row buffers in NeuPIMs~\cite{neupims}.
Our cycle-accurate DRAM simulation via Ramulator 2.0 ensures that the control-path throughput is accurately measured and that no timing violations occur between DRAM and PIM commands~\cite{samsung_hbm2pim}.
We adopt a performance model for GPUs from Duplex~\cite{duplex}, assuming experts are executed via grouped GEMM on GPUs.
We configure the GPU to match the NVIDIA B200 GPU as shown in Table~\ref{tab:memconfig_b200}.
All experiments assume multi-GPU systems with HBM-PIM stacks where each HBM-PIM die is accessible only through its attached GPU.
Although \moepim is evaluated by simulating multi-GPU systems with HBM-PIM stacks in which each HBM-PIM die is accessible only through its attached GPU, its core ideas also apply to the global interconnect scenarios considered in prior work, where any XPU in the system can directly access any PIM device~\cite{neupims, attacc, papi, pimoe}.

\textbf{Models: }
We evaluate \moepim on GPT-OSS-120B (\textit{GPT-OSS}), Qwen3.5-397B-A17B (\textit{Qwen3.5}), and Qwen3-30B-A3B (\textit{Qwen3}).
These models also represent the recent LLM trends discussed in Section~\ref{sec:llmtrends}. 
\textit{GPT-OSS} and \textit{Qwen3} activate four experts out of 128, while \textit{Qwen3.5} activates ten experts out of 512 with one shared expert. 
Furthermore, \textit{Qwen3.5} and \textit{GPT-OSS} reflect the trend toward greater sparsity in state-of-the-art MoE models, but 
along different dimensions. 
Earlier MoE models typically activate eight experts out of 128 or 256 total experts~\cite{deepseekR1,deepseekV3,qwen3,ernie}.
\textit{Qwen3.5} increases the act-ratio by increasing the total number of experts, whereas \textit{GPT-OSS} does so by lowering the ratio of activated to total experts.
We evaluate a range of batch sizes ($B$) up to 256 to satisfy common Service Level Objectives (SLOs)~\cite{paise,papi,duplex,pimba,neupims,specpim,glm45fp8_hf}.
We use four B200 GPUs for \textit{GPT-OSS}, eight B200 GPUs for \textit{Qwen3.5}, and a single B200 GPU for \textit{Qwen3}, where all GPUs are attached to their HBM-PIM stacks.

\textbf{Datasets: }
We collect real expert distributions by running the MoE models on GPUs,
using real-world request traces from the HH-RLHF~\cite{hh-rlhf} and MATH-500~\cite{math500} datasets.
HH-RLHF contains human preference comparisons for long, open-ended dialogue sequences across diverse topics~\cite{hh-rlhf}. 
MATH-500 is a benchmark of competition-style math problems with formal statements and proof-like reasoning~\cite{math500}. 
Unless stated otherwise, activations and parameters are stored in bfloat16.

\textbf{Comparison Methodology:}
We compare \moepim against three methods that statically determine which experts are executed on PIM: \textit{NoExp}, \textit{AllExp}, and \textit{PIMoE}, each reproducing a state-of-the-art in- and near-memory computing technique for LLM inference~\cite{papi,duplex,neupims,attacc,paise,stratum,pimoe}.
Each method incorporates the same non-MoE optimizations, such as offloading attention operations to PIM~\cite{papi,duplex,neupims,attacc,paise}, distributing KV cache across PIM channels~\cite{neupims}, and maintaining separate paths for DRAM and PIM commands~\cite{duplex,neupims}.
All methods share identical hardware configurations and differ only in how expert computations are scheduled between GPUs and their HBM-PIM stacks.

\textbf{\textit{NoExp}} denotes a method where only attention operations are offloaded to PIM and all experts are executed on GPUs~\cite{neupims,paise}.
\textit{NoExp} with attention operations offloaded to PIM is the dominant and widely studied PIM acceleration strategy, providing a consistent point of comparison across expert offloading techniques.
\textbf{\textit{AllExp}} executes all experts on PIM~\cite{papi,stratum}.
Although Stratum~\cite{stratum} targets monolithic 3D-Stackable DRAM rather than HBM-PIM, we reproduce its scheduling policy, which performs all expert computations in the decode phase using near-memory processing.
PAPI~\cite{papi} similarly explains that PIM with a larger number of processing units can efficiently handle expert computations.
\textbf{\textit{PIMoE}} assumes that all experts are assigned to PIM first and moves the most popular expert from the busiest PIM channel to the GPU until the GPU execution time becomes larger than the PIM execution time~\cite{pimoe}.

Since a request yields one prefill phase followed by multiple decode phases, approximately 90\% of stages in continuous batching correspond to decoding-only phases~\cite{duplex, samsungPIM_hcs23}.
Moreover, in modern inference systems, the prefill and decode phases are often executed on separate resources, a practice known as prefill-decode disaggregation~\cite{splitwise, distserve}.
As a result, efficiently processing batches composed solely of decode-phase requests has become critical for practical deployments.
Therefore, we mainly evaluate \moepim in the case where all requests in a batch are in the decode phase.
Scenarios that mix prefill and decode phases are evaluated in Section~\ref{sec:colocated-pd}.

\subsection{Results}

\label{sec:main-result}
\textbf{Evaluation Metrics for the Pareto Curve:}
Various parallelization strategies, including data parallelism, tensor parallelism, context parallelism, and expert parallelism, can be combined depending on whether a system targets high throughput or low latency.
Therefore, rather than evaluating \moepim using either throughput or latency alone,
we use a Pareto curve defined by interactivity (generated tokens per second per user) and throughput per GPU (generated tokens per second per GPU).
Higher interactivity corresponds to lower latency, and points in the upper-right region represent efficient system performance.

\textbf{Throughput and Interactivity Improvements by \moepim: } Figure~\ref{fig:main-result} illustrates the throughput (tokens per second) and interactivity of \moepim compared to prior methods across various batch sizes on \textit{GPT-OSS} and \textit{Qwen3.5}. 
\moepim consistently outperforms all baselines, delivering substantial gains in both total system throughput and per-user interactivity. 
Crucially, \moepim demonstrates superior system performance and scalability compared to the optimized \textit{PIMoE} baseline. 
For small batch sizes, \textit{PIMoE} and \textit{AllExp} achieve performance comparable to \moepim, since most experts are memory-bound and assigning all experts to PIM is effective in this setting.

However, their performance starts to degrade with larger batch sizes, as the proportion of compute-bound experts increases. 
\moepim adaptively assigns such compute-bound experts to GPUs, achieving higher speedups than \textit{PIMoE} and \textit{AllExp} at larger batch sizes. 
On \textit{Qwen3.5}, \moepim delivers up to a $26\%$ ($1.26\times$) improvement over \textit{PIMoE} in both throughput and interactivity at $B=256$.
Similarly, on \textit{GPT-OSS}, it yields a steady $11\%$-$17\%$ ($1.11\times$-$1.17\times$) throughput gain across moderate to large batch sizes ($B \ge 32$), with interactivity improvements reaching up to $1.25\times$.
Ultimately, \moepim is the only approach that smoothly scales peak throughput at high loads while strictly satisfying interactivity service-level agreements (SLAs).

\textbf{Analysis of Pareto Curve: } Figure~\ref{fig:main-result} reveals distinct performance trajectories for the baselines and \moepim. 
\textit{NoExp}, which relies heavily on GPUs, exhibits an L-shaped curve, indicating suboptimal scaling with early throughput saturation.
Conversely, executing all experts on the PIM (\textit{AllExp}) yields an almost flat horizontal trajectory; it maintains interactivity, but throughput fails to scale beyond $B=32$.
\moepim and \textit{PIMoE} achieve a curve closer to the ideal inverted-L frontier. 

This behavior is strongly correlated with how the expert distribution evolves as the batch size increases.
At highly constrained batch sizes ($B \le 16$), the extreme sparsity of models like \textit{Qwen3.5} reduces most expert computations to memory-bound operations. 
In this regime, confining execution to the PIM is strictly optimal, allowing \moepim and \textit{PIMoE} to match \textit{AllExp}'s high performance. 
However, as the batch size grows past $16$ and approaches $32$, the workload generates popular experts with higher arithmetic intensity. 
The key to achieving a curve closer to the ideal inverted-L trajectory is recognizing this transition point. 
By selectively offloading these hot experts to the GPU at $B \ge 32$, \moepim surpasses the throughput ceiling at which \textit{AllExp} saturates.

\textbf{Architectural Advantages of \moepim: } While both \moepim and \textit{PIMoE} achieve curves closer to the ideal inverted-L trajectory compared to other baselines, \moepim achieves significantly higher peak throughput (a $\sim1.26\times$ improvement for \textit{Qwen3.5}) by addressing two critical architectural blind spots. 
First, \textit{PIMoE} balances expert placement by evaluating isolated expert execution times but ignores attention computation on the PIM, whose cost grows rapidly at larger batch sizes. 
Second, in real-world MoE inference, an optimal schedule can only be achieved when inter-GPU communication, compute, and memory demands are considered together. 
\textit{PIMoE} overlooks the overhead of communication across GPUs and overloads the PIM, extending PIM execution beyond the latency of network transfers and thereby limiting overall throughput. 

In contrast, \moepim explicitly incorporates both network communication costs and attention overheads into its scheduling policy.
As a result, \moepim achieves higher throughput and interactivity than prior work in both multi-GPU (\texttt{Qwen3.5} and \textit{GPT-OSS} in \Cref{fig:main-result}) and single-GPU settings (\textit{Qwen3} in \Cref{fig:main-result}).
In single-GPU settings, \moepim achieves up to a $1.6\times$ improvement in throughput and interactivity over \textit{PIMoE} by accounting for the attention overheads in the scheduling policy.
In multi-GPU settings, \moepim reduces PIM latency below the communication threshold by accounting for network communication costs and attention overheads, unlocking an additional $1.3\times$ improvement in throughput and interactivity over \textit{PIMoE} on \textit{Qwen3.5} and \textit{GPT-OSS}.
To conclude, \moepim effectively shifts the Pareto frontier in both single- and multi-GPU setups, guaranteeing that the system processes exponentially more tokens globally without penalizing the responsiveness of individual user requests.

\label{sec:pimchannels}
\begin{figure}[t]
    \centering
    \includegraphics[width=\columnwidth]{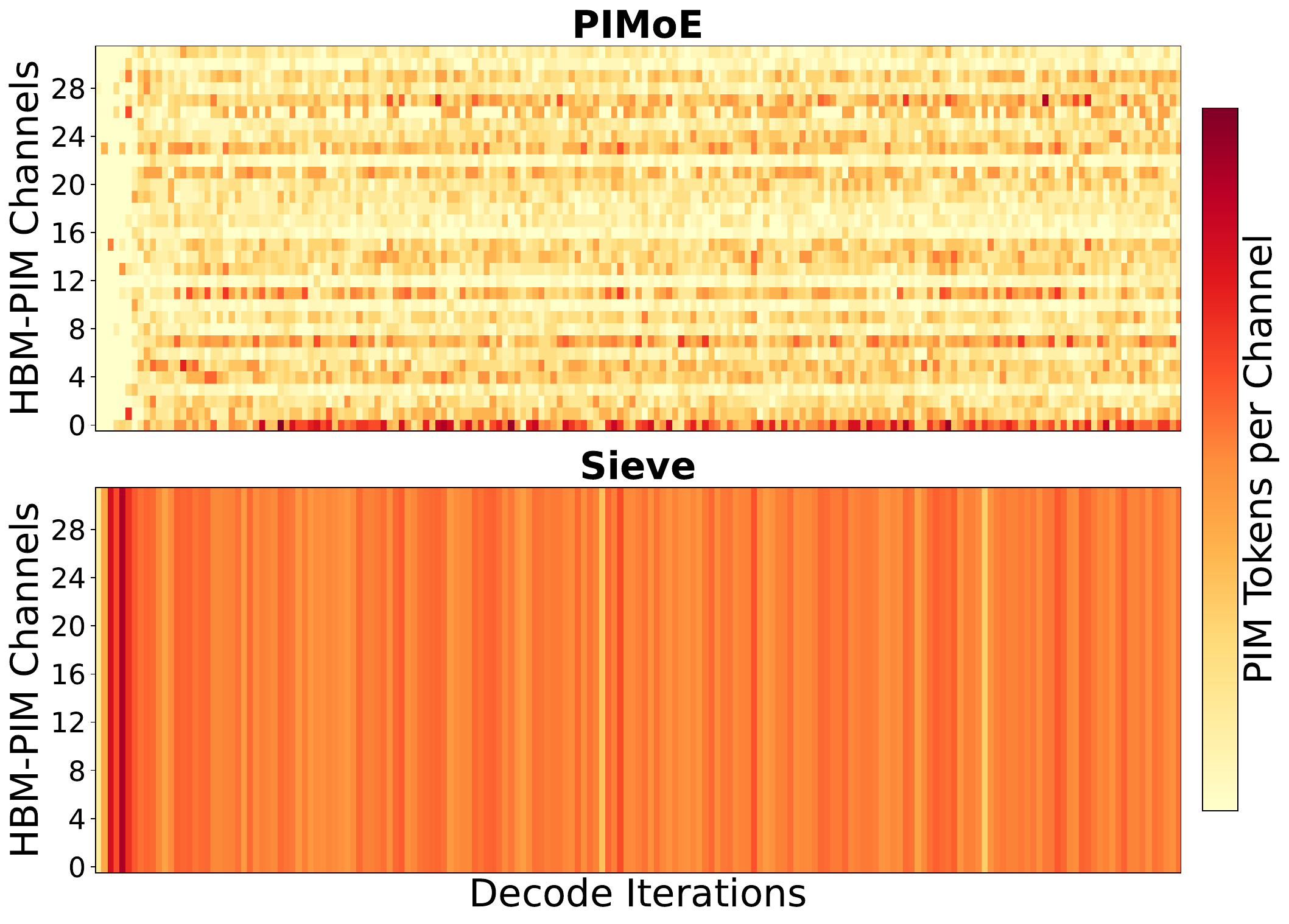}
    \Description{Heatmap showing PIM channel utilization across 4 B200 GPUs when running GPT-OSS at batch size 64.}
    \caption{Utilization of PIM channels when running \textit{GPT-OSS} with 4 B200 GPUs equipped with HBM-PIM stacks. A darker color means a PIM channel has higher utilization. 
    }
    \label{fig:pimchannels}
\end{figure}
\textbf{Parallelism Across PIM Channels: }
As discussed in Section~\ref{sec:pimfriendly}, \moepim employs tensor parallelism across PIM channels within a GPU, while \textit{PIMoE} uses expert parallelism.
The arithmetic intensity disparity across experts leads to significant fluctuations in PIM channel utilization in \textit{PIMoE}, as illustrated in Figure~\ref{fig:pimchannels}; certain channels experience heavy workloads while others remain underutilized.
Furthermore, since commercial PIM architectures lack direct data transfer between PIM channels~\cite{psyncpim}, the applicability of the Expert Parallelism Load Balancer across PIM channels in \textit{PIMoE} is limited, making it difficult to mitigate this hardware-level inefficiency.
However, since \moepim utilizes tensor parallelism across PIM channels, every channel is uniformly utilized as long as memory-bound experts are executed on PIM.
This approach effectively decouples PIM channel utilization from the expert distribution.

\subsection{Colocated Prefill-Decode Requests}
\label{sec:colocated-pd}

Previous results demonstrate that \moepim is a promising solution for efficient MoE serving within Prefill-Decode (PD) disaggregation, although PD disaggregation may not always represent practical deployments.
To address this, we evaluate \moepim under colocated PD in \textit{Qwen3}, where each batch contains both prefill-phase and decode-phase requests.
State-of-the-art LLM frameworks such as vLLM~\cite{vllm} limit batches to at most two prefill-phase requests because queries often exceed 1024 tokens.
To stress-test \moepim, however, we construct extreme cases with up to eight prefill-phase requests per batch for $B \ge 64$, and up to two prefill-phase requests per batch for $B \le 32$.

\begin{figure}[t]
    \centering
    \includegraphics[width=\columnwidth]{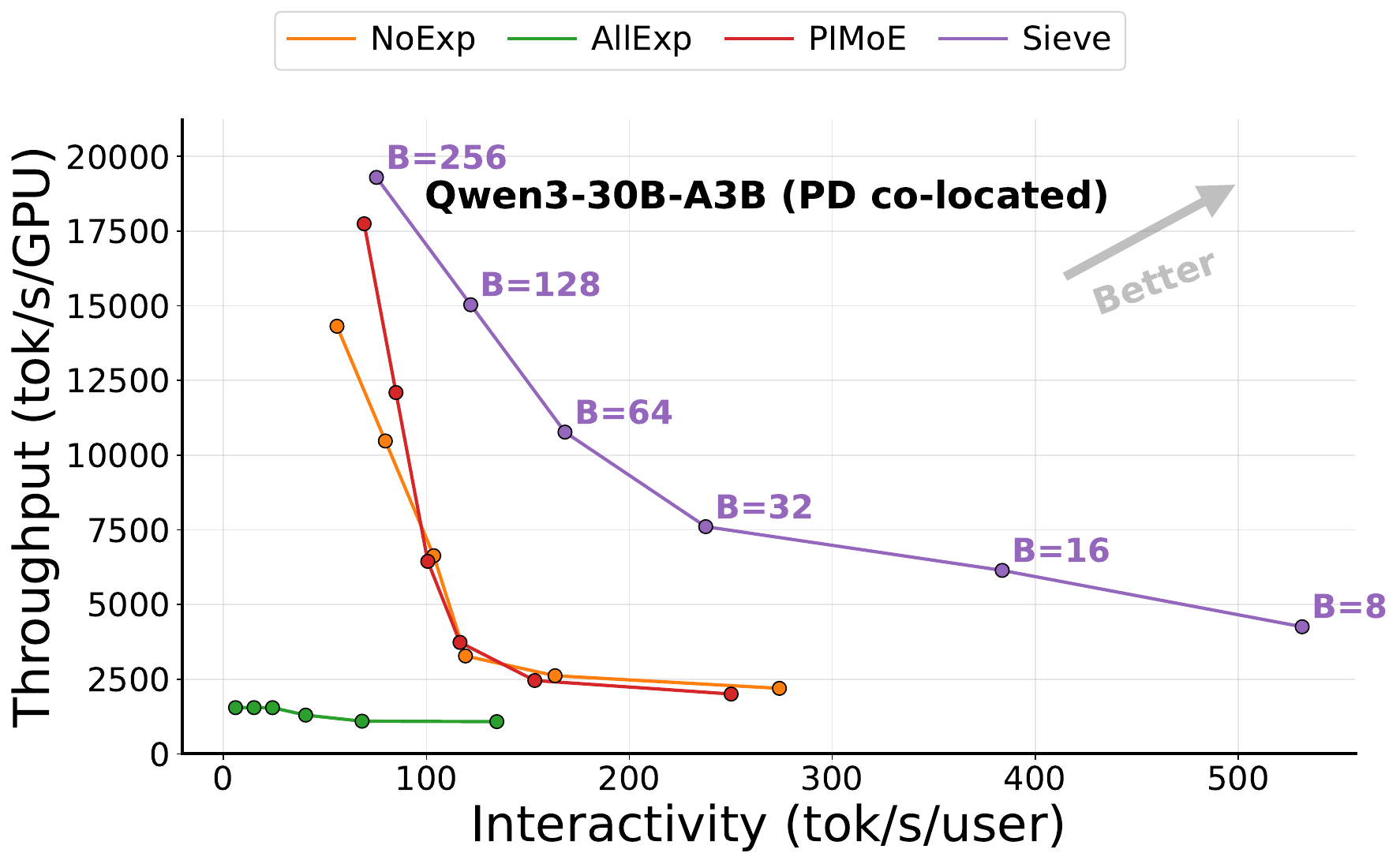}
    \Description{Pareto curve comparing throughput and interactivity of Sieve, NoExp, AllExp, and PIMoE for Qwen-3 under colocated prefill-decode requests.}
    \caption{Evaluation of throughput and interactivity achieved by \moepim, \textit{NoExp}, \textit{AllExp}, and \textit{PIMoE} for \textit{Qwen-3} under colocated prefill-decode requests.}
    \label{fig:copd}
\end{figure}

As shown in Figure~\ref{fig:copd}, \moepim consistently outperforms all prior methods under colocated prefill-decode requests.
Even in extreme cases where eight prefill-phase requests coexist in a batch, \moepim achieves $2.4\times$ ($B=16$) and $2.3\times$ ($B=32$) speedups compared to \textit{NoExp}, respectively.
In contrast, both \textit{AllExp} and \textit{PIMoE} exhibit noticeable slowdowns as the number of prefill-phase requests increases, often performing worse than \textit{NoExp}.
This degradation occurs because a higher fraction of prefill requests increases the probability of executing compute-bound GEMM experts on PIM, a situation that \moepim effectively avoids.
As there are eight prefill requests at $B \ge 64$, the number of compute-bound experts becomes larger.
Moreover, \moepim achieves greater throughput gains at $B \ge 64$ compared to $B \le 32$ as shown in Figure~\ref{fig:copd}.
Consequently, \moepim can efficiently support MoE serving in both colocated prefill-decode and disaggregated prefill-decode scenarios by maintaining balanced utilization between PIM and GPU.

%% file: tab_memconfig_b200gpu.tex
\begin{table}[t]
  \centering
  \small
  \setlength{\tabcolsep}{2.3pt}
  \renewcommand{\arraystretch}{1.05}
  \begin{tabular}{|c|c||c|c|}
    \hline
    \multicolumn{2}{|c||}{\textbf{GPU Configuration (B200 GPU)}} &
    \multicolumn{2}{c|}{\textbf{HBM-PIM Configuration}} \\
    \hline
    FP16 throughput     & 2,250\,TFLOPS     & HBM generation        & HBM3E \\
    \hline
    HBM-PIM bandwidth       & 8.0\,TB/s         & Pin rate              & 8.0\,Gbps \\
    \hline
    HBM-PIM stacks      & 8                 & Pseudo-channels/stack & 32 \\
    \hline
    HBM-PIM capacity$^\dagger$ & 96\,GB     & Banks/pseudo-channel  & 24 \\
    \hline
    NVLink BW (per dir.) & 900\,GB/s        & Page size             & 1\,KB \\
    \hline
    NVLink latency       & 0.8\,$\mu$s      & Compute density       & 1\,op/byte \\
    \hline
    \multicolumn{4}{|c|}{\textbf{HBM3E Timing Parameters (cycles @ 8.0\,Gbps, tCK $\approx$ 0.50\,ns)}} \\
    \hline
    \multicolumn{4}{|c|}{\makecell[c]{
      tRCD = 28,\; tRP = 28,\; tRAS = 68,\; tRC = 96,\; tCL = 28,\\
      tWR = 32,\; tCCD\_S = 2,\; tCCD\_L = 4,\; tRRD\_S = 6,\; tRRD\_L = 6,\\
      tFAW = 12,\; tREFI = 3,900\,ns,\; tRFC = 400\,ns}} \\
    \hline
  \end{tabular}
  \caption{Specification of the evaluated B200 multi-GPU system with HBM-PIM.
  $^\dagger$PIM processing units sacrifice 50\% of HBM capacity (192$\to$96\,GB per GPU).}
  \label{tab:memconfig_b200}
\end{table}

%% file: sections/8_related.tex
\section{Related Work}
Recent research has extensively explored PIM to mitigate the memory wall in LLM inference \cite{neupims, attacc, paise, papi}. 
A dominant direction is to offload memory-bound operators, such as attention computation, to PIM while keeping compute-bound operators such as dense FFNs on the xPU \cite{neupims, attacc, paise, papi}.
These studies show the effectiveness of heterogeneous xPU--PIM execution for dense transformers, but largely assume that the offloaded portion is known a priori from the model structure or memory layout.

A second line of work considers MoE-aware heterogeneous execution. 
Prior studies observe that MoE layers exhibit varying arithmetic intensity and advocate heterogeneous execution across xPU and PIM \cite{duplex, pimoe}. However, prior approaches rely on static or coarse-grained placement policies that do not fully capture the highly imbalanced expert distribution observed in recent sparse MoE models \cite{duplex, pimoe}.

\moepim{} is complementary to orthogonal efforts that improve the PIM substrate or target bottlenecks, such as long-context attention or KV-cache management \cite{pimphony, orches, longsight, repa}. 
These approaches can be combined with \moepim{} because they do not address the same problem of dynamically partitioning MoE expert computations between GPU and PIM based on the runtime expert distribution. 
In contrast, \moepim{} targets modern and future MoE serving scenarios and provides a lightweight runtime scheduler that maps unpopular, memory-bound experts to PIM while keeping popular experts on the xPU.

%% file: sections/9_conclusion.tex
\section{Conclusion}

This paper presents a scheduler for MoE models running on multi-GPU systems with PIM, accounting for both the arithmetic intensity disparity and inter-GPU communication incurred by expert parallelism.
Furthermore, we propose \moepim, a framework for the proposed scheduler, which accelerates modern MoE models on PIM-enabled systems by efficiently coordinating operations on GPUs and PIM.
Our evaluations on state-of-the-art MoE models demonstrate throughput and interactivity gains over prior PIM systems for MoE.

%% file: main.bbl

\begin{thebibliography}{57}


\ifx \showCODEN    \undefined \def \showCODEN     #1{\unskip}     \fi
\ifx \showDOI      \undefined \def \showDOI       #1{#1}\fi
\ifx \showISBNx    \undefined \def \showISBNx     #1{\unskip}     \fi
\ifx \showISBNxiii \undefined \def \showISBNxiii  #1{\unskip}     \fi
\ifx \showISSN     \undefined \def \showISSN      #1{\unskip}     \fi
\ifx \showLCCN     \undefined \def \showLCCN      #1{\unskip}     \fi
\ifx \shownote     \undefined \def \shownote      #1{#1}          \fi
\ifx \showarticletitle \undefined \def \showarticletitle #1{#1}   \fi
\ifx \showURL      \undefined \def \showURL       {\relax}        \fi
\providecommand\bibfield[2]{#2}
\providecommand\bibinfo[2]{#2}
\providecommand\natexlab[1]{#1}
\providecommand\showeprint[2][]{arXiv:#2}

\bibitem[glm({[n.\,d.]})]%
        {glm45fp8_hf}
 \bibinfo{year}{[n.\,d.]}\natexlab{}.
\newblock \bibinfo{title}{GLM-4.5-FP8 Model Card}.
\newblock
  \bibinfo{howpublished}{\url{https://huggingface.co/zai-org/GLM-4.5-FP8}}.
\newblock
\newblock
\shownote{Accessed: 2025-11-15}.


\bibitem[Agarwal et~al\mbox{.}(2025)]%
        {gptoss}
\bibfield{author}{\bibinfo{person}{Sandhini Agarwal}, \bibinfo{person}{Lama
  Ahmad}, \bibinfo{person}{Jason Ai}, \bibinfo{person}{Sam Altman},
  \bibinfo{person}{Andy Applebaum}, \bibinfo{person}{Edwin Arbus},
  \bibinfo{person}{Rahul~K Arora}, \bibinfo{person}{Yu Bai},
  \bibinfo{person}{Bowen Baker}, \bibinfo{person}{Haiming Bao},
  {et~al\mbox{.}}} \bibinfo{year}{2025}\natexlab{}.
\newblock \showarticletitle{gpt-oss-120b \& gpt-oss-20b model card}.
\newblock \bibinfo{journal}{\emph{arXiv preprint arXiv:2508.10925}}
  (\bibinfo{year}{2025}).
\newblock


\bibitem[{Artificial Analysis}(2025)]%
        {artificialanalysis_2025}
\bibfield{author}{\bibinfo{person}{{Artificial Analysis}}.}
  \bibinfo{year}{2025}\natexlab{}.
\newblock \bibinfo{title}{Artificial Analysis}.
\newblock \bibinfo{howpublished}{\url{https://artificialanalysis.ai}}.
\newblock
\newblock
\shownote{Accessed: 2025-11-03}.


\bibitem[Baek et~al\mbox{.}(2024)]%
        {psyncpim}
\bibfield{author}{\bibinfo{person}{Daehyeon Baek}, \bibinfo{person}{Soojin
  Hwang}, {and} \bibinfo{person}{Jaehyuk Huh}.}
  \bibinfo{year}{2024}\natexlab{}.
\newblock \showarticletitle{pSyncPIM: Partially Synchronous Execution of Sparse
  Matrix Operations for All-Bank PIM Architectures}. In
  \bibinfo{booktitle}{\emph{2024 ACM/IEEE 51st Annual International Symposium
  on Computer Architecture (ISCA)}}. IEEE, \bibinfo{pages}{354--367}.
\newblock


\bibitem[Baidu-ERNIE-Team(2025)]%
        {ernie}
\bibfield{author}{\bibinfo{person}{Baidu-ERNIE-Team}.}
  \bibinfo{year}{2025}\natexlab{}.
\newblock \bibinfo{title}{ERNIE 4.5 Technical Report}.
\newblock
\newblock
\urldef\tempurl%
\url{https://yiyan.baidu.com/blog/publication/ERNIE_Technical_Report.pdf}
\showURL{%
\tempurl}


\bibitem[Brown et~al\mbox{.}(2020)]%
        {gpt3}
\bibfield{author}{\bibinfo{person}{Tom Brown}, \bibinfo{person}{Benjamin Mann},
  \bibinfo{person}{Nick Ryder}, \bibinfo{person}{Melanie Subbiah},
  \bibinfo{person}{Jared~D Kaplan}, \bibinfo{person}{Prafulla Dhariwal},
  \bibinfo{person}{Arvind Neelakantan}, \bibinfo{person}{Pranav Shyam},
  \bibinfo{person}{Girish Sastry}, \bibinfo{person}{Amanda Askell},
  {et~al\mbox{.}}} \bibinfo{year}{2020}\natexlab{}.
\newblock \showarticletitle{Language models are few-shot learners}.
\newblock \bibinfo{journal}{\emph{Advances in neural information processing
  systems}}  \bibinfo{volume}{33} (\bibinfo{year}{2020}),
  \bibinfo{pages}{1877--1901}.
\newblock


\bibitem[Chen et~al\mbox{.}(2025)]%
        {ktransformers}
\bibfield{author}{\bibinfo{person}{Hongtao Chen}, \bibinfo{person}{Weiyu Xie},
  \bibinfo{person}{Boxin Zhang}, \bibinfo{person}{Jingqi Tang},
  \bibinfo{person}{Jiahao Wang}, \bibinfo{person}{Jianwei Dong},
  \bibinfo{person}{Shaoyuan Chen}, \bibinfo{person}{Ziwei Yuan},
  \bibinfo{person}{Chen Lin}, \bibinfo{person}{Chengyu Qiu},
  \bibinfo{person}{Yuening Zhu}, \bibinfo{person}{Qingliang Ou},
  \bibinfo{person}{Jiaqi Liao}, \bibinfo{person}{Xianglin Chen},
  \bibinfo{person}{Zhiyuan Ai}, \bibinfo{person}{Yongwei Wu}, {and}
  \bibinfo{person}{Mingxing Zhang}.} \bibinfo{year}{2025}\natexlab{}.
\newblock \showarticletitle{KTransformers: Unleashing the Full Potential of
  CPU/GPU Hybrid Inference for MoE Models}. In
  \bibinfo{booktitle}{\emph{Proceedings of the ACM SIGOPS 31st Symposium on
  Operating Systems Principles}}.
\newblock


\bibitem[Chen et~al\mbox{.}(2014)]%
        {dadiannao}
\bibfield{author}{\bibinfo{person}{Yunji Chen}, \bibinfo{person}{Tao Luo},
  \bibinfo{person}{Shaoli Liu}, \bibinfo{person}{Shijin Zhang},
  \bibinfo{person}{Liqiang He}, \bibinfo{person}{Jia Wang},
  \bibinfo{person}{Ling Li}, \bibinfo{person}{Tianshi Chen},
  \bibinfo{person}{Zhiwei Xu}, \bibinfo{person}{Ninghui Sun}, {et~al\mbox{.}}}
  \bibinfo{year}{2014}\natexlab{}.
\newblock \showarticletitle{Dadiannao: A machine-learning supercomputer}. In
  \bibinfo{booktitle}{\emph{2014 47th Annual IEEE/ACM International Symposium
  on Microarchitecture}}. IEEE, \bibinfo{pages}{609--622}.
\newblock


\bibitem[Chen et~al\mbox{.}(2016)]%
        {eyeriss}
\bibfield{author}{\bibinfo{person}{Yu-Hsin Chen}, \bibinfo{person}{Joel Emer},
  {and} \bibinfo{person}{Vivienne Sze}.} \bibinfo{year}{2016}\natexlab{}.
\newblock \showarticletitle{Eyeriss: A spatial architecture for
  energy-efficient dataflow for convolutional neural networks}.
\newblock \bibinfo{journal}{\emph{ACM SIGARCH computer architecture news}}
  \bibinfo{volume}{44}, \bibinfo{number}{3} (\bibinfo{year}{2016}),
  \bibinfo{pages}{367--379}.
\newblock


\bibitem[Comanici et~al\mbox{.}(2025)]%
        {gemini2.5}
\bibfield{author}{\bibinfo{person}{Gheorghe Comanici}, \bibinfo{person}{Eric
  Bieber}, \bibinfo{person}{Mike Schaekermann}, \bibinfo{person}{Ice Pasupat},
  \bibinfo{person}{Noveen Sachdeva}, \bibinfo{person}{Inderjit Dhillon},
  \bibinfo{person}{Marcel Blistein}, \bibinfo{person}{Ori Ram},
  \bibinfo{person}{Dan Zhang}, \bibinfo{person}{Evan Rosen}, {et~al\mbox{.}}}
  \bibinfo{year}{2025}\natexlab{}.
\newblock \showarticletitle{Gemini 2.5: Pushing the frontier with advanced
  reasoning, multimodality, long context, and next generation agentic
  capabilities}.
\newblock \bibinfo{journal}{\emph{arXiv preprint arXiv:2507.06261}}
  (\bibinfo{year}{2025}).
\newblock


\bibitem[Dai et~al\mbox{.}(2024)]%
        {deepseekmoe}
\bibfield{author}{\bibinfo{person}{Damai Dai}, \bibinfo{person}{Chengqi Deng},
  \bibinfo{person}{Chenggang Zhao}, \bibinfo{person}{R.~X. Xu},
  \bibinfo{person}{Huazuo Gao}, \bibinfo{person}{Deli Chen},
  \bibinfo{person}{Jiashi Li}, \bibinfo{person}{Wangding Zeng},
  \bibinfo{person}{Xingkai Yu}, \bibinfo{person}{Y. Wu},
  \bibinfo{person}{Zhenda Xie}, \bibinfo{person}{Y.~K. Li},
  \bibinfo{person}{Panpan Huang}, \bibinfo{person}{Fuli Luo},
  \bibinfo{person}{Chong Ruan}, \bibinfo{person}{Zhifang Sui}, {and}
  \bibinfo{person}{Wenfeng Liang}.} \bibinfo{year}{2024}\natexlab{}.
\newblock \bibinfo{title}{DeepSeekMoE: Towards Ultimate Expert Specialization
  in Mixture-of-Experts Language Models}.
\newblock
\newblock
\showeprint[arxiv]{2401.06066}~[cs.CL]
\urldef\tempurl%
\url{https://arxiv.org/abs/2401.06066}
\showURL{%
\tempurl}


\bibitem[Dally(2023)]%
        {dally2023hardware}
\bibfield{author}{\bibinfo{person}{Bill Dally}.}
  \bibinfo{year}{2023}\natexlab{}.
\newblock \showarticletitle{Hardware for deep learning}. In
  \bibinfo{booktitle}{\emph{2023 IEEE Hot Chips 35 Symposium (HCS)}}. IEEE
  Computer Society, \bibinfo{pages}{1--58}.
\newblock


\bibitem[Devlin et~al\mbox{.}(2019)]%
        {bert}
\bibfield{author}{\bibinfo{person}{Jacob Devlin}, \bibinfo{person}{Ming-Wei
  Chang}, \bibinfo{person}{Kenton Lee}, {and} \bibinfo{person}{Kristina
  Toutanova}.} \bibinfo{year}{2019}\natexlab{}.
\newblock \showarticletitle{Bert: Pre-training of deep bidirectional
  transformers for language understanding}. In
  \bibinfo{booktitle}{\emph{Proceedings of the 2019 conference of the North
  American chapter of the association for computational linguistics: human
  language technologies, volume 1 (long and short papers)}}.
  \bibinfo{pages}{4171--4186}.
\newblock


\bibitem[Du et~al\mbox{.}(2022)]%
        {du2022glam}
\bibfield{author}{\bibinfo{person}{Nan Du}, \bibinfo{person}{Yanping Huang},
  \bibinfo{person}{Andrew~M Dai}, \bibinfo{person}{Simon Tong},
  \bibinfo{person}{Dmitry Lepikhin}, \bibinfo{person}{Yuanzhong Xu},
  \bibinfo{person}{Maxim Krikun}, \bibinfo{person}{Yanqi Zhou},
  \bibinfo{person}{Adams~Wei Yu}, \bibinfo{person}{Orhan Firat},
  {et~al\mbox{.}}} \bibinfo{year}{2022}\natexlab{}.
\newblock \showarticletitle{Glam: Efficient scaling of language models with
  mixture-of-experts}. In \bibinfo{booktitle}{\emph{International conference on
  machine learning}}. PMLR, \bibinfo{pages}{5547--5569}.
\newblock


\bibitem[Fedus et~al\mbox{.}(2022)]%
        {fedus2022switch}
\bibfield{author}{\bibinfo{person}{William Fedus}, \bibinfo{person}{Barret
  Zoph}, {and} \bibinfo{person}{Noam Shazeer}.}
  \bibinfo{year}{2022}\natexlab{}.
\newblock \showarticletitle{Switch transformers: Scaling to trillion parameter
  models with simple and efficient sparsity}.
\newblock \bibinfo{journal}{\emph{Journal of Machine Learning Research}}
  \bibinfo{volume}{23}, \bibinfo{number}{120} (\bibinfo{year}{2022}),
  \bibinfo{pages}{1--39}.
\newblock


\bibitem[Ganguli et~al\mbox{.}(2022)]%
        {hh-rlhf}
\bibfield{author}{\bibinfo{person}{Deep Ganguli}, \bibinfo{person}{Liane
  Lovitt}, \bibinfo{person}{Jackson Kernion}, \bibinfo{person}{Amanda Askell},
  \bibinfo{person}{Yuntao Bai}, \bibinfo{person}{Saurav Kadavath},
  \bibinfo{person}{Ben Mann}, \bibinfo{person}{Ethan Perez},
  \bibinfo{person}{Nicholas Schiefer}, \bibinfo{person}{Kamal Ndousse},
  \bibinfo{person}{Andy Jones}, \bibinfo{person}{Sam Bowman},
  \bibinfo{person}{Anna Chen}, \bibinfo{person}{Tom Conerly},
  \bibinfo{person}{Nova DasSarma}, \bibinfo{person}{Dawn Drain},
  \bibinfo{person}{Nelson Elhage}, \bibinfo{person}{Sheer El-Showk},
  \bibinfo{person}{Stanislav Fort}, \bibinfo{person}{Zac Hatfield-Dodds},
  \bibinfo{person}{Tom Henighan}, \bibinfo{person}{Danny Hernandez},
  \bibinfo{person}{Tristan Hume}, \bibinfo{person}{Josh Jacobson},
  \bibinfo{person}{Scott Johnston}, \bibinfo{person}{Shauna Kravec},
  \bibinfo{person}{Catherine Olsson}, \bibinfo{person}{Sam Ringer},
  \bibinfo{person}{Eli Tran-Johnson}, \bibinfo{person}{Dario Amodei},
  \bibinfo{person}{Tom Brown}, \bibinfo{person}{Nicholas Joseph},
  \bibinfo{person}{Sam McCandlish}, \bibinfo{person}{Chris Olah},
  \bibinfo{person}{Jared Kaplan}, {and} \bibinfo{person}{Jack Clark}.}
  \bibinfo{year}{2022}\natexlab{}.
\newblock \bibinfo{title}{Red Teaming Language Models to Reduce Harms: Methods,
  Scaling Behaviors, and Lessons Learned}.
\newblock
\newblock
\showeprint[arxiv]{2209.07858}~[cs.CL]
\urldef\tempurl%
\url{https://arxiv.org/abs/2209.07858}
\showURL{%
\tempurl}


\bibitem[Gholami et~al\mbox{.}(2024)]%
        {aiandmemorywall}
\bibfield{author}{\bibinfo{person}{Amir Gholami}, \bibinfo{person}{Zhewei Yao},
  \bibinfo{person}{Sehoon Kim}, \bibinfo{person}{Coleman Hooper},
  \bibinfo{person}{Michael~W Mahoney}, {and} \bibinfo{person}{Kurt Keutzer}.}
  \bibinfo{year}{2024}\natexlab{}.
\newblock \showarticletitle{Ai and memory wall}.
\newblock \bibinfo{journal}{\emph{IEEE Micro}} \bibinfo{volume}{44},
  \bibinfo{number}{3} (\bibinfo{year}{2024}), \bibinfo{pages}{33--39}.
\newblock


\bibitem[Gu et~al\mbox{.}(2025)]%
        {pimisallyouneed}
\bibfield{author}{\bibinfo{person}{Yufeng Gu}, \bibinfo{person}{Alireza
  Khadem}, \bibinfo{person}{Sumanth Umesh}, \bibinfo{person}{Ning Liang},
  \bibinfo{person}{Xavier Servot}, \bibinfo{person}{Onur Mutlu},
  \bibinfo{person}{Ravi Iyer}, {and} \bibinfo{person}{Reetuparna Das}.}
  \bibinfo{year}{2025}\natexlab{}.
\newblock \showarticletitle{PIM is all you need: A CXL-enabled GPU-free system
  for large language model inference}. In \bibinfo{booktitle}{\emph{Proceedings
  of the 30th ACM International Conference on Architectural Support for
  Programming Languages and Operating Systems, Volume 2}}.
  \bibinfo{pages}{862--881}.
\newblock


\bibitem[Guo et~al\mbox{.}(2025)]%
        {deepseekR1}
\bibfield{author}{\bibinfo{person}{Daya Guo}, \bibinfo{person}{Dejian Yang},
  \bibinfo{person}{Haowei Zhang}, \bibinfo{person}{Junxiao Song},
  \bibinfo{person}{Ruoyu Zhang}, \bibinfo{person}{Runxin Xu},
  \bibinfo{person}{Qihao Zhu}, \bibinfo{person}{Shirong Ma},
  \bibinfo{person}{Peiyi Wang}, \bibinfo{person}{Xiao Bi}, {et~al\mbox{.}}}
  \bibinfo{year}{2025}\natexlab{}.
\newblock \showarticletitle{Deepseek-r1: Incentivizing reasoning capability in
  llms via reinforcement learning}.
\newblock \bibinfo{journal}{\emph{arXiv preprint arXiv:2501.12948}}
  (\bibinfo{year}{2025}).
\newblock


\bibitem[He et~al\mbox{.}(2020)]%
        {newton}
\bibfield{author}{\bibinfo{person}{Mingxuan He}, \bibinfo{person}{Choungki
  Song}, \bibinfo{person}{Ilkon Kim}, \bibinfo{person}{Chunseok Jeong},
  \bibinfo{person}{Seho Kim}, \bibinfo{person}{Il Park},
  \bibinfo{person}{Mithuna Thottethodi}, {and} \bibinfo{person}{TN
  Vijaykumar}.} \bibinfo{year}{2020}\natexlab{}.
\newblock \showarticletitle{Newton: A DRAM-maker’s accelerator-in-memory
  (AiM) architecture for machine learning}. In \bibinfo{booktitle}{\emph{2020
  53rd Annual IEEE/ACM International Symposium on Microarchitecture (MICRO)}}.
  IEEE, \bibinfo{pages}{372--385}.
\newblock


\bibitem[He et~al\mbox{.}(2025)]%
        {papi}
\bibfield{author}{\bibinfo{person}{Yintao He}, \bibinfo{person}{Haiyu Mao},
  \bibinfo{person}{Christina Giannoula}, \bibinfo{person}{Mohammad
  Sadrosadati}, \bibinfo{person}{Juan G{\'o}mez-Luna}, \bibinfo{person}{Huawei
  Li}, \bibinfo{person}{Xiaowei Li}, \bibinfo{person}{Ying Wang}, {and}
  \bibinfo{person}{Onur Mutlu}.} \bibinfo{year}{2025}\natexlab{}.
\newblock \showarticletitle{Papi: Exploiting dynamic parallelism in large
  language model decoding with a processing-in-memory-enabled computing
  system}. In \bibinfo{booktitle}{\emph{Proceedings of the 30th ACM
  International Conference on Architectural Support for Programming Languages
  and Operating Systems, Volume 2}}. \bibinfo{pages}{766--782}.
\newblock


\bibitem[Heo et~al\mbox{.}(2024)]%
        {neupims}
\bibfield{author}{\bibinfo{person}{Guseul Heo}, \bibinfo{person}{Sangyeop Lee},
  \bibinfo{person}{Jaehong Cho}, \bibinfo{person}{Hyunmin Choi},
  \bibinfo{person}{Sanghyeon Lee}, \bibinfo{person}{Hyungkyu Ham},
  \bibinfo{person}{Gwangsun Kim}, \bibinfo{person}{Divya Mahajan}, {and}
  \bibinfo{person}{Jongse Park}.} \bibinfo{year}{2024}\natexlab{}.
\newblock \showarticletitle{Neupims: Npu-pim heterogeneous acceleration for
  batched llm inferencing}. In \bibinfo{booktitle}{\emph{Proceedings of the
  29th ACM International Conference on Architectural Support for Programming
  Languages and Operating Systems, Volume 3}}. \bibinfo{pages}{722--737}.
\newblock


\bibitem[Hong et~al\mbox{.}(2026)]%
        {repa}
\bibfield{author}{\bibinfo{person}{Yang Hong}, \bibinfo{person}{Junlong Yang},
  \bibinfo{person}{Bo Peng}, {and} \bibinfo{person}{Jianguo Yao}.}
  \bibinfo{year}{2026}\natexlab{}.
\newblock \showarticletitle{REPA: Re configurable P IM for the Joint A
  cceleration of KV Cache Offloading and Processing}. In
  \bibinfo{booktitle}{\emph{Proceedings of the 31st ACM International
  Conference on Architectural Support for Programming Languages and Operating
  Systems, Volume 2}}. \bibinfo{pages}{1622--1639}.
\newblock


\bibitem[Jiang et~al\mbox{.}(2024)]%
        {mixtral}
\bibfield{author}{\bibinfo{person}{Albert~Q. Jiang}, \bibinfo{person}{Alexandre
  Sablayrolles}, \bibinfo{person}{Antoine Roux}, \bibinfo{person}{Arthur
  Mensch}, \bibinfo{person}{Blanche Savary}, \bibinfo{person}{Chris Bamford},
  \bibinfo{person}{Devendra~Singh Chaplot}, \bibinfo{person}{Diego de~las
  Casas}, \bibinfo{person}{Emma~Bou Hanna}, \bibinfo{person}{Florian Bressand},
  \bibinfo{person}{Gianna Lengyel}, \bibinfo{person}{Guillaume Bour},
  \bibinfo{person}{Guillaume Lample}, \bibinfo{person}{Lélio~Renard Lavaud},
  \bibinfo{person}{Lucile Saulnier}, \bibinfo{person}{Marie-Anne Lachaux},
  \bibinfo{person}{Pierre Stock}, \bibinfo{person}{Sandeep Subramanian},
  \bibinfo{person}{Sophia Yang}, \bibinfo{person}{Szymon Antoniak},
  \bibinfo{person}{Teven~Le Scao}, \bibinfo{person}{Théophile Gervet},
  \bibinfo{person}{Thibaut Lavril}, \bibinfo{person}{Thomas Wang},
  \bibinfo{person}{Timothée Lacroix}, {and} \bibinfo{person}{William~El
  Sayed}.} \bibinfo{year}{2024}\natexlab{}.
\newblock \bibinfo{title}{Mixtral of Experts}.
\newblock
\newblock
\showeprint[arxiv]{2401.04088}~[cs.LG]
\urldef\tempurl%
\url{https://arxiv.org/abs/2401.04088}
\showURL{%
\tempurl}


\bibitem[Jouppi et~al\mbox{.}(2017)]%
        {tpuv1}
\bibfield{author}{\bibinfo{person}{Norman~P Jouppi}, \bibinfo{person}{Cliff
  Young}, \bibinfo{person}{Nishant Patil}, \bibinfo{person}{David Patterson},
  \bibinfo{person}{Gaurav Agrawal}, \bibinfo{person}{Raminder Bajwa},
  \bibinfo{person}{Sarah Bates}, \bibinfo{person}{Suresh Bhatia},
  \bibinfo{person}{Nan Boden}, \bibinfo{person}{Al Borchers}, {et~al\mbox{.}}}
  \bibinfo{year}{2017}\natexlab{}.
\newblock \showarticletitle{In-datacenter performance analysis of a tensor
  processing unit}. In \bibinfo{booktitle}{\emph{Proceedings of the 44th annual
  international symposium on computer architecture}}. \bibinfo{pages}{1--12}.
\newblock


\bibitem[Kim et~al\mbox{.}(2022)]%
        {aquabolt}
\bibfield{author}{\bibinfo{person}{Jin~Hyun Kim}, \bibinfo{person}{Shin-Haeng
  Kang}, \bibinfo{person}{Sukhan Lee}, \bibinfo{person}{Hyeonsu Kim},
  \bibinfo{person}{Yuhwan Ro}, \bibinfo{person}{Seungwon Lee},
  \bibinfo{person}{David Wang}, \bibinfo{person}{Jihyun Choi},
  \bibinfo{person}{Jinin So}, \bibinfo{person}{YeonGon Cho}, {et~al\mbox{.}}}
  \bibinfo{year}{2022}\natexlab{}.
\newblock \showarticletitle{Aquabolt-XL HBM2-PIM, LPDDR5-PIM with in-memory
  processing, and AXDIMM with acceleration buffer}.
\newblock \bibinfo{journal}{\emph{IEEE Micro}} \bibinfo{volume}{42},
  \bibinfo{number}{3} (\bibinfo{year}{2022}), \bibinfo{pages}{20--30}.
\newblock


\bibitem[Kim et~al\mbox{.}(2023)]%
        {samsungPIM_hcs23}
\bibfield{author}{\bibinfo{person}{Jin~Hyun Kim}, \bibinfo{person}{Yuhwan Ro},
  \bibinfo{person}{Jinin So}, \bibinfo{person}{Sukhan Lee},
  \bibinfo{person}{Shin-haeng Kang}, \bibinfo{person}{YeonGon Cho},
  \bibinfo{person}{Hyeonsu Kim}, \bibinfo{person}{Byeongho Kim},
  \bibinfo{person}{Kyungsoo Kim}, \bibinfo{person}{Sangsoo Park},
  {et~al\mbox{.}}} \bibinfo{year}{2023}\natexlab{}.
\newblock \showarticletitle{Samsung pim/pnm for transfmer based ai: Energy
  efficiency on pim/pnm cluster}. In \bibinfo{booktitle}{\emph{2023 IEEE Hot
  Chips 35 Symposium (HCS)}}. IEEE Computer Society, \bibinfo{pages}{1--31}.
\newblock


\bibitem[Kim et~al\mbox{.}(2025)]%
        {pimba}
\bibfield{author}{\bibinfo{person}{Wonung Kim}, \bibinfo{person}{Yubin Lee},
  \bibinfo{person}{Yoonsung Kim}, \bibinfo{person}{Jinwoo Hwang},
  \bibinfo{person}{Seongryong Oh}, \bibinfo{person}{Jiyong Jung},
  \bibinfo{person}{Aziz Huseynov}, \bibinfo{person}{Woong~Gyu Park},
  \bibinfo{person}{Chang~Hyun Park}, \bibinfo{person}{Divya Mahajan},
  {et~al\mbox{.}}} \bibinfo{year}{2025}\natexlab{}.
\newblock \showarticletitle{Pimba: A Processing-in-Memory Acceleration for
  Post-Transformer Large Language Model Serving}.
\newblock \bibinfo{journal}{\emph{arXiv preprint arXiv:2507.10178}}
  (\bibinfo{year}{2025}).
\newblock


\bibitem[Kwon et~al\mbox{.}(2026)]%
        {pimphony}
\bibfield{author}{\bibinfo{person}{Hyucksung Kwon}, \bibinfo{person}{Kyungmo
  Koo}, \bibinfo{person}{Janghyeon Kim}, \bibinfo{person}{Woongkyu Lee},
  \bibinfo{person}{Minjae Lee}, \bibinfo{person}{Gyeonggeun Jung},
  \bibinfo{person}{Hyungdeok Lee}, \bibinfo{person}{Yousub Jung},
  \bibinfo{person}{Jaehan Park}, \bibinfo{person}{Yosub Song}, {et~al\mbox{.}}}
  \bibinfo{year}{2026}\natexlab{}.
\newblock \showarticletitle{PIMphony: Overcoming Bandwidth and Capacity
  Inefficiency in PIM-Based Long-Context LLM Inference System}. In
  \bibinfo{booktitle}{\emph{2026 IEEE International Symposium on High
  Performance Computer Architecture (HPCA)}}. IEEE, \bibinfo{pages}{1--21}.
\newblock


\bibitem[Kwon et~al\mbox{.}(2023)]%
        {vllm}
\bibfield{author}{\bibinfo{person}{Woosuk Kwon}, \bibinfo{person}{Zhuohan Li},
  \bibinfo{person}{Siyuan Zhuang}, \bibinfo{person}{Ying Sheng},
  \bibinfo{person}{Lianmin Zheng}, \bibinfo{person}{Cody~Hao Yu},
  \bibinfo{person}{Joseph~E. Gonzalez}, \bibinfo{person}{Hao Zhang}, {and}
  \bibinfo{person}{Ion Stoica}.} \bibinfo{year}{2023}\natexlab{}.
\newblock \showarticletitle{Efficient Memory Management for Large Language
  Model Serving with PagedAttention}. In \bibinfo{booktitle}{\emph{Proceedings
  of the ACM SIGOPS 29th Symposium on Operating Systems Principles}}.
\newblock


\bibitem[Kwon et~al\mbox{.}(2022)]%
        {gddr6aim}
\bibfield{author}{\bibinfo{person}{Yongkee Kwon}, \bibinfo{person}{Kornijcuk
  Vladimir}, \bibinfo{person}{Nahsung Kim}, \bibinfo{person}{Woojae Shin},
  \bibinfo{person}{Jongsoon Won}, \bibinfo{person}{Minkyu Lee},
  \bibinfo{person}{Hyunha Joo}, \bibinfo{person}{Haerang Choi},
  \bibinfo{person}{Guhyun Kim}, \bibinfo{person}{Byeongju An}, {et~al\mbox{.}}}
  \bibinfo{year}{2022}\natexlab{}.
\newblock \showarticletitle{System architecture and software stack for
  GDDR6-AiM}. In \bibinfo{booktitle}{\emph{2022 IEEE Hot Chips 34 Symposium
  (HCS)}}. IEEE, \bibinfo{pages}{1--25}.
\newblock


\bibitem[Lee et~al\mbox{.}(2025)]%
        {paise}
\bibfield{author}{\bibinfo{person}{Hyojung Lee}, \bibinfo{person}{Daehyeon
  Baek}, \bibinfo{person}{Jimyoung Son}, \bibinfo{person}{Jieun Choi},
  \bibinfo{person}{Kihyo Moon}, {and} \bibinfo{person}{Minsung Jang}.}
  \bibinfo{year}{2025}\natexlab{}.
\newblock \showarticletitle{PAISE: PIM-Accelerated Inference Scheduling Engine
  for Transformer-based LLM}. In \bibinfo{booktitle}{\emph{2025 IEEE
  International Symposium on High Performance Computer Architecture (HPCA)}}.
  IEEE, \bibinfo{pages}{1707--1719}.
\newblock


\bibitem[Lee et~al\mbox{.}(2021)]%
        {samsung_hbm2pim}
\bibfield{author}{\bibinfo{person}{Sukhan Lee}, \bibinfo{person}{Shin-haeng
  Kang}, \bibinfo{person}{Jaehoon Lee}, \bibinfo{person}{Hyeonsu Kim},
  \bibinfo{person}{Eojin Lee}, \bibinfo{person}{Seungwoo Seo},
  \bibinfo{person}{Hosang Yoon}, \bibinfo{person}{Seungwon Lee},
  \bibinfo{person}{Kyounghwan Lim}, \bibinfo{person}{Hyunsung Shin},
  {et~al\mbox{.}}} \bibinfo{year}{2021}\natexlab{}.
\newblock \showarticletitle{Hardware architecture and software stack for PIM
  based on commercial DRAM technology: Industrial product}. In
  \bibinfo{booktitle}{\emph{2021 ACM/IEEE 48th Annual International Symposium
  on Computer Architecture (ISCA)}}. IEEE, \bibinfo{pages}{43--56}.
\newblock


\bibitem[Lee et~al\mbox{.}(2022)]%
        {hynix_pim_isscc22}
\bibfield{author}{\bibinfo{person}{Seongju Lee}, \bibinfo{person}{Kyuyoung
  Kim}, \bibinfo{person}{Sanghoon Oh}, \bibinfo{person}{Joonhong Park},
  \bibinfo{person}{Gimoon Hong}, \bibinfo{person}{Dongyoon Ka},
  \bibinfo{person}{Kyudong Hwang}, \bibinfo{person}{Jeongje Park},
  \bibinfo{person}{Kyeongpil Kang}, \bibinfo{person}{Jungyeon Kim},
  {et~al\mbox{.}}} \bibinfo{year}{2022}\natexlab{}.
\newblock \showarticletitle{A 1ynm 1.25 V 8Gb, 16Gb/s/pin GDDR6-based
  accelerator-in-memory supporting 1TFLOPS MAC operation and various activation
  functions for deep-learning applications}. In \bibinfo{booktitle}{\emph{2022
  IEEE International Solid-State Circuits Conference (ISSCC)}},
  Vol.~\bibinfo{volume}{65}. IEEE, \bibinfo{pages}{1--3}.
\newblock


\bibitem[Lepikhin et~al\mbox{.}(2020)]%
        {lepikhin2020gshard}
\bibfield{author}{\bibinfo{person}{Dmitry Lepikhin},
  \bibinfo{person}{HyoukJoong Lee}, \bibinfo{person}{Yuanzhong Xu},
  \bibinfo{person}{Dehao Chen}, \bibinfo{person}{Orhan Firat},
  \bibinfo{person}{Yanping Huang}, \bibinfo{person}{Maxim Krikun},
  \bibinfo{person}{Noam Shazeer}, {and} \bibinfo{person}{Zhifeng Chen}.}
  \bibinfo{year}{2020}\natexlab{}.
\newblock \showarticletitle{Gshard: Scaling giant models with conditional
  computation and automatic sharding}.
\newblock \bibinfo{journal}{\emph{arXiv preprint arXiv:2006.16668}}
  (\bibinfo{year}{2020}).
\newblock


\bibitem[Li et~al\mbox{.}(2024)]%
        {specpim}
\bibfield{author}{\bibinfo{person}{Cong Li}, \bibinfo{person}{Zhe Zhou},
  \bibinfo{person}{Size Zheng}, \bibinfo{person}{Jiaxi Zhang},
  \bibinfo{person}{Yun Liang}, {and} \bibinfo{person}{Guangyu Sun}.}
  \bibinfo{year}{2024}\natexlab{}.
\newblock \showarticletitle{Specpim: Accelerating speculative inference on
  pim-enabled system via architecture-dataflow co-exploration}. In
  \bibinfo{booktitle}{\emph{Proceedings of the 29th ACM International
  Conference on Architectural Support for Programming Languages and Operating
  Systems, Volume 3}}. \bibinfo{pages}{950--965}.
\newblock


\bibitem[Li et~al\mbox{.}(2025)]%
        {orches}
\bibfield{author}{\bibinfo{person}{Sixu Li}, \bibinfo{person}{Yuzhou Chen},
  \bibinfo{person}{Chaojian Li}, \bibinfo{person}{Yonggan Fu},
  \bibinfo{person}{Zheng Wang}, \bibinfo{person}{Zhongzhi Yu},
  \bibinfo{person}{Haoran You}, \bibinfo{person}{Zhifan Ye},
  \bibinfo{person}{Wei Zhou}, \bibinfo{person}{Yongan Zhang}, {et~al\mbox{.}}}
  \bibinfo{year}{2025}\natexlab{}.
\newblock \showarticletitle{ORCHES: Orchestrated Test-Time-Compute-based LLM
  Reasoning on Collaborative GPU-PIM HEterogeneous System}. In
  \bibinfo{booktitle}{\emph{Proceedings of the 58th IEEE/ACM International
  Symposium on Microarchitecture}}. \bibinfo{pages}{476--489}.
\newblock


\bibitem[Lightman et~al\mbox{.}(2023)]%
        {math500}
\bibfield{author}{\bibinfo{person}{Hunter Lightman}, \bibinfo{person}{Vineet
  Kosaraju}, \bibinfo{person}{Yuri Burda}, \bibinfo{person}{Harrison Edwards},
  \bibinfo{person}{Bowen Baker}, \bibinfo{person}{Teddy Lee},
  \bibinfo{person}{Jan Leike}, \bibinfo{person}{John Schulman},
  \bibinfo{person}{Ilya Sutskever}, {and} \bibinfo{person}{Karl Cobbe}.}
  \bibinfo{year}{2023}\natexlab{}.
\newblock \showarticletitle{Let's verify step by step}. In
  \bibinfo{booktitle}{\emph{The Twelfth International Conference on Learning
  Representations}}.
\newblock


\bibitem[Liu et~al\mbox{.}(2024)]%
        {deepseekV3}
\bibfield{author}{\bibinfo{person}{Aixin Liu}, \bibinfo{person}{Bei Feng},
  \bibinfo{person}{Bing Xue}, \bibinfo{person}{Bingxuan Wang},
  \bibinfo{person}{Bochao Wu}, \bibinfo{person}{Chengda Lu},
  \bibinfo{person}{Chenggang Zhao}, \bibinfo{person}{Chengqi Deng},
  \bibinfo{person}{Chenyu Zhang}, \bibinfo{person}{Chong Ruan},
  {et~al\mbox{.}}} \bibinfo{year}{2024}\natexlab{}.
\newblock \showarticletitle{Deepseek-v3 technical report}.
\newblock \bibinfo{journal}{\emph{arXiv preprint arXiv:2412.19437}}
  (\bibinfo{year}{2024}).
\newblock


\bibitem[Luo et~al\mbox{.}(2023)]%
        {ramulator2}
\bibfield{author}{\bibinfo{person}{Haocong Luo}, \bibinfo{person}{Yahya~Can
  Tu\u{g}rul}, \bibinfo{person}{F.~Nisa Bostancı}, \bibinfo{person}{Ataberk
  Olgun}, \bibinfo{person}{A.~Giray Ya\u{g}l{\i}k\c{c}{\i}},
  \bibinfo{person}{}, {and} \bibinfo{person}{Onur Mutlu}.}
  \bibinfo{year}{2023}\natexlab{}.
\newblock \bibinfo{title}{{Ramulator 2.0: A Modern, Modular, and Extensible
  DRAM Simulator}}.
\newblock
\newblock


\bibitem[{Meta AI}(2025)]%
        {meta2025llama4}
\bibfield{author}{\bibinfo{person}{{Meta AI}}.}
  \bibinfo{year}{2025}\natexlab{}.
\newblock \bibinfo{title}{The Llama 4 Herd: The Beginning of a New Era of
  Natively Multimodal Models}.
\newblock
\newblock
\urldef\tempurl%
\url{https://ai.meta.com/blog/llama-4-multimodal-intelligence/}
\showURL{%
\tempurl}


\bibitem[{NVIDIA}(2025)]%
        {nvidia-b200-gpu}
\bibfield{author}{\bibinfo{person}{{NVIDIA}}.} \bibinfo{year}{2025}\natexlab{}.
\newblock \bibinfo{title}{NVIDIA DGX B200 Datasheet}.
\newblock
\newblock
\newblock
\shownote{https://resources.nvidia.com/en-us-dgx-systems/dgx-b200-datasheet}.


\bibitem[Pan et~al\mbox{.}(2025)]%
        {stratum}
\bibfield{author}{\bibinfo{person}{Yue Pan}, \bibinfo{person}{Zihan Xia},
  \bibinfo{person}{Po-Kai Hsu}, \bibinfo{person}{Lanxiang Hu},
  \bibinfo{person}{Hyungyo Kim}, \bibinfo{person}{Janak Sharda},
  \bibinfo{person}{Minxuan Zhou}, \bibinfo{person}{Nam~Sung Kim},
  \bibinfo{person}{Shimeng Yu}, \bibinfo{person}{Tajana Rosing},
  {et~al\mbox{.}}} \bibinfo{year}{2025}\natexlab{}.
\newblock \showarticletitle{Stratum: System-Hardware Co-Design with Tiered
  Monolithic 3D-Stackable DRAM for Efficient MoE Serving}.
\newblock \bibinfo{journal}{\emph{arXiv preprint arXiv:2510.05245}}
  (\bibinfo{year}{2025}).
\newblock


\bibitem[Park et~al\mbox{.}(2024)]%
        {attacc}
\bibfield{author}{\bibinfo{person}{Jaehyun Park}, \bibinfo{person}{Jaewan
  Choi}, \bibinfo{person}{Kwanhee Kyung}, \bibinfo{person}{Michael~Jaemin Kim},
  \bibinfo{person}{Yongsuk Kwon}, \bibinfo{person}{Nam~Sung Kim}, {and}
  \bibinfo{person}{Jung~Ho Ahn}.} \bibinfo{year}{2024}\natexlab{}.
\newblock \showarticletitle{Attacc! unleashing the power of pim for batched
  transformer-based generative model inference}. In
  \bibinfo{booktitle}{\emph{Proceedings of the 29th ACM International
  Conference on Architectural Support for Programming Languages and Operating
  Systems, Volume 2}}. \bibinfo{pages}{103--119}.
\newblock


\bibitem[Patel et~al\mbox{.}(2024)]%
        {splitwise}
\bibfield{author}{\bibinfo{person}{Pratyush Patel}, \bibinfo{person}{Esha
  Choukse}, \bibinfo{person}{Chaojie Zhang}, \bibinfo{person}{Aashaka Shah},
  \bibinfo{person}{{\'I}{\~n}igo Goiri}, \bibinfo{person}{Saeed Maleki}, {and}
  \bibinfo{person}{Ricardo Bianchini}.} \bibinfo{year}{2024}\natexlab{}.
\newblock \showarticletitle{Splitwise: Efficient generative llm inference using
  phase splitting}. In \bibinfo{booktitle}{\emph{2024 ACM/IEEE 51st Annual
  International Symposium on Computer Architecture (ISCA)}}. IEEE,
  \bibinfo{pages}{118--132}.
\newblock


\bibitem[Quinn et~al\mbox{.}(2025)]%
        {longsight}
\bibfield{author}{\bibinfo{person}{Derrick Quinn}, \bibinfo{person}{E~Ezgi
  Y{\"u}cel}, \bibinfo{person}{Jinkwon Kim}, \bibinfo{person}{Jos{\'e}~F
  Mart{\'\i}nez}, {and} \bibinfo{person}{Mohammad Alian}.}
  \bibinfo{year}{2025}\natexlab{}.
\newblock \showarticletitle{LongSight: Compute-Enabled Memory to Accelerate
  Large-Context LLMs via Sparse Attention}. In
  \bibinfo{booktitle}{\emph{Proceedings of the 58th IEEE/ACM International
  Symposium on Microarchitecture}}. \bibinfo{pages}{34--48}.
\newblock


\bibitem[Shazeer et~al\mbox{.}(2017)]%
        {shazeer2017outrageously}
\bibfield{author}{\bibinfo{person}{Noam Shazeer}, \bibinfo{person}{Azalia
  Mirhoseini}, \bibinfo{person}{Krzysztof Maziarz}, \bibinfo{person}{Andy
  Davis}, \bibinfo{person}{Quoc Le}, \bibinfo{person}{Geoffrey Hinton}, {and}
  \bibinfo{person}{Jeff Dean}.} \bibinfo{year}{2017}\natexlab{}.
\newblock \showarticletitle{Outrageously large neural networks: The
  sparsely-gated mixture-of-experts layer}.
\newblock \bibinfo{journal}{\emph{arXiv preprint arXiv:1701.06538}}
  (\bibinfo{year}{2017}).
\newblock


\bibitem[Team(2025)]%
        {glm4.5}
\bibfield{author}{\bibinfo{person}{GLM-4.5 Team}.}
  \bibinfo{year}{2025}\natexlab{}.
\newblock \bibinfo{title}{GLM-4.5: Agentic, Reasoning, and Coding (ARC)
  Foundation Models}.
\newblock
\newblock
\urldef\tempurl%
\url{https://arxiv.org/abs/2508.06471}
\showURL{%
\tempurl}


\bibitem[Team et~al\mbox{.}(2025)]%
        {kimiK2}
\bibfield{author}{\bibinfo{person}{Kimi Team}, \bibinfo{person}{Yifan Bai},
  \bibinfo{person}{Yiping Bao}, \bibinfo{person}{Guanduo Chen},
  \bibinfo{person}{Jiahao Chen}, \bibinfo{person}{Ningxin Chen},
  \bibinfo{person}{Ruijue Chen}, \bibinfo{person}{Yanru Chen},
  \bibinfo{person}{Yuankun Chen}, \bibinfo{person}{Yutian Chen},
  {et~al\mbox{.}}} \bibinfo{year}{2025}\natexlab{}.
\newblock \showarticletitle{Kimi K2: Open Agentic Intelligence}.
\newblock \bibinfo{journal}{\emph{arXiv preprint arXiv:2507.20534}}
  (\bibinfo{year}{2025}).
\newblock


\bibitem[Wolf et~al\mbox{.}(2019)]%
        {huggingfacetransformers}
\bibfield{author}{\bibinfo{person}{Thomas Wolf}, \bibinfo{person}{Lysandre
  Debut}, \bibinfo{person}{Victor Sanh}, \bibinfo{person}{Julien Chaumond},
  \bibinfo{person}{Clement Delangue}, \bibinfo{person}{Anthony Moi},
  \bibinfo{person}{Pierric Cistac}, \bibinfo{person}{Tim Rault},
  \bibinfo{person}{R{\'e}mi Louf}, \bibinfo{person}{Morgan Funtowicz},
  {et~al\mbox{.}}} \bibinfo{year}{2019}\natexlab{}.
\newblock \showarticletitle{Huggingface's transformers: State-of-the-art
  natural language processing}.
\newblock \bibinfo{journal}{\emph{arXiv preprint arXiv:1910.03771}}
  (\bibinfo{year}{2019}).
\newblock


\bibitem[Wu et~al\mbox{.}(2025)]%
        {pimoe}
\bibfield{author}{\bibinfo{person}{Lizhou Wu}, \bibinfo{person}{Haozhe Zhu},
  \bibinfo{person}{Siqi He}, \bibinfo{person}{Xuanda Lin},
  \bibinfo{person}{Xiaoyang Zeng}, {and} \bibinfo{person}{Chixiao Chen}.}
  \bibinfo{year}{2025}\natexlab{}.
\newblock \showarticletitle{PIMoE: Towards efficient MoE transformer deployment
  on NPU-PIM system through throttle-aware task offloading}. In
  \bibinfo{booktitle}{\emph{2025 62nd ACM/IEEE Design Automation Conference
  (DAC)}}. IEEE, \bibinfo{pages}{1--7}.
\newblock


\bibitem[Yang et~al\mbox{.}(2025)]%
        {qwen3}
\bibfield{author}{\bibinfo{person}{An Yang}, \bibinfo{person}{Anfeng Li},
  \bibinfo{person}{Baosong Yang}, \bibinfo{person}{Beichen Zhang},
  \bibinfo{person}{Binyuan Hui}, \bibinfo{person}{Bo Zheng},
  \bibinfo{person}{Bowen Yu}, \bibinfo{person}{Chang Gao},
  \bibinfo{person}{Chengen Huang}, \bibinfo{person}{Chenxu Lv},
  {et~al\mbox{.}}} \bibinfo{year}{2025}\natexlab{}.
\newblock \showarticletitle{Qwen3 technical report}.
\newblock \bibinfo{journal}{\emph{arXiv preprint arXiv:2505.09388}}
  (\bibinfo{year}{2025}).
\newblock


\bibitem[Yun et~al\mbox{.}(2024)]%
        {duplex}
\bibfield{author}{\bibinfo{person}{Sungmin Yun}, \bibinfo{person}{Kwanhee
  Kyung}, \bibinfo{person}{Juhwan Cho}, \bibinfo{person}{Jaewan Choi},
  \bibinfo{person}{Jongmin Kim}, \bibinfo{person}{Byeongho Kim},
  \bibinfo{person}{Sukhan Lee}, \bibinfo{person}{Kyomin Sohn}, {and}
  \bibinfo{person}{Jung~Ho Ahn}.} \bibinfo{year}{2024}\natexlab{}.
\newblock \showarticletitle{Duplex: A device for large language models with
  mixture of experts, grouped query attention, and continuous batching}. In
  \bibinfo{booktitle}{\emph{2024 57th IEEE/ACM International Symposium on
  Microarchitecture (MICRO)}}. IEEE, \bibinfo{pages}{1429--1443}.
\newblock


\bibitem[Zhang et~al\mbox{.}(2025)]%
        {comet}
\bibfield{author}{\bibinfo{person}{Shulai Zhang}, \bibinfo{person}{Ningxin
  Zheng}, \bibinfo{person}{Haibin Lin}, \bibinfo{person}{Ziheng Jiang},
  \bibinfo{person}{Wenlei Bao}, \bibinfo{person}{Chengquan Jiang},
  \bibinfo{person}{Qi Hou}, \bibinfo{person}{Weihao Cui}, \bibinfo{person}{Size
  Zheng}, \bibinfo{person}{Li-Wen Chang}, {et~al\mbox{.}}}
  \bibinfo{year}{2025}\natexlab{}.
\newblock \showarticletitle{Comet: Fine-grained computation-communication
  overlapping for mixture-of-experts}.
\newblock \bibinfo{journal}{\emph{Proceedings of Machine Learning and Systems}}
   \bibinfo{volume}{7} (\bibinfo{year}{2025}).
\newblock


\bibitem[Zheng et~al\mbox{.}(2024)]%
        {sglang}
\bibfield{author}{\bibinfo{person}{Lianmin Zheng}, \bibinfo{person}{Liangsheng
  Yin}, \bibinfo{person}{Zhiqiang Xie}, \bibinfo{person}{Chuyue~Livia Sun},
  \bibinfo{person}{Jeff Huang}, \bibinfo{person}{Cody~Hao Yu},
  \bibinfo{person}{Shiyi Cao}, \bibinfo{person}{Christos Kozyrakis},
  \bibinfo{person}{Ion Stoica}, \bibinfo{person}{Joseph~E Gonzalez},
  {et~al\mbox{.}}} \bibinfo{year}{2024}\natexlab{}.
\newblock \showarticletitle{Sglang: Efficient execution of structured language
  model programs}.
\newblock \bibinfo{journal}{\emph{Advances in neural information processing
  systems}}  \bibinfo{volume}{37} (\bibinfo{year}{2024}),
  \bibinfo{pages}{62557--62583}.
\newblock


\bibitem[Zhong et~al\mbox{.}(2024)]%
        {distserve}
\bibfield{author}{\bibinfo{person}{Yinmin Zhong}, \bibinfo{person}{Shengyu
  Liu}, \bibinfo{person}{Junda Chen}, \bibinfo{person}{Jianbo Hu},
  \bibinfo{person}{Yibo Zhu}, \bibinfo{person}{Xuanzhe Liu},
  \bibinfo{person}{Xin Jin}, {and} \bibinfo{person}{Hao Zhang}.}
  \bibinfo{year}{2024}\natexlab{}.
\newblock \showarticletitle{DistServe: disaggregating prefill and decoding for
  goodput-optimized large language model serving}. In
  \bibinfo{booktitle}{\emph{Proceedings of the 18th USENIX Conference on
  Operating Systems Design and Implementation}} (Santa Clara, CA, USA)
  \emph{(\bibinfo{series}{OSDI'24})}. \bibinfo{publisher}{USENIX Association},
  \bibinfo{address}{USA}, Article \bibinfo{articleno}{11},
  \bibinfo{numpages}{18}~pages.
\newblock
\showISBNx{978-1-939133-40-3}


\bibitem[Zhu et~al\mbox{.}(2025)]%
        {megascale-infer}
\bibfield{author}{\bibinfo{person}{Ruidong Zhu}, \bibinfo{person}{Ziheng
  Jiang}, \bibinfo{person}{Chao Jin}, \bibinfo{person}{Peng Wu},
  \bibinfo{person}{Cesar~A Stuardo}, \bibinfo{person}{Dongyang Wang},
  \bibinfo{person}{Xinlei Zhang}, \bibinfo{person}{Huaping Zhou},
  \bibinfo{person}{Haoran Wei}, \bibinfo{person}{Yang Cheng}, {et~al\mbox{.}}}
  \bibinfo{year}{2025}\natexlab{}.
\newblock \showarticletitle{Megascale-infer: Serving mixture-of-experts at
  scale with disaggregated expert parallelism}.
\newblock \bibinfo{journal}{\emph{arXiv preprint arXiv:2504.02263}}
  (\bibinfo{year}{2025}).
\newblock


\end{thebibliography}
